\documentclass[letterpaper, english, aps, prl, reprint, twocolumn, superscriptaddress]{revtex4-2}
\usepackage[T1]{fontenc}
\usepackage{color}
\usepackage{amsmath}
\usepackage{amssymb}
\usepackage{graphicx}
\usepackage{multirow}
\usepackage{booktabs}
\usepackage{capt-of}
\usepackage[latin9]{inputenc}
\usepackage{babel}
\usepackage{calc}
\usepackage{float}
\makeatletter

\usepackage[bookmarks=true,colorlinks,linkcolor=black,anchorcolor=green,citecolor=blue,urlcolor=blue]{hyperref}

\renewcommand{\fnum@figure}{FIG. \thefigure}
\renewcommand{\fnum@table}{TABLE. \thetable}

\makeatother

\begin{document}
	\title{Stable Higher-Order Topological Dirac Semimetals with $\mathbb{Z}_2$ Monopole Charge in Alternating-twisted Multilayer Graphenes and beyond}
	
	\author{Shifeng Qian}
	\affiliation{Centre for Quantum Physics, Key Laboratory of Advanced Optoelectronic Quantum Architecture and Measurement (MOE), School of Physics, Beijing Institute of Technology, Beijing, 100081, China}

	\author{Yongpan Li}
	\affiliation{Centre for Quantum Physics, Key Laboratory of Advanced Optoelectronic Quantum Architecture and Measurement (MOE), School of Physics, Beijing Institute of Technology, Beijing, 100081, China}

	\author{Cheng-Cheng Liu}
	\email{ccliu@bit.edu.cn}
	\affiliation{Centre for Quantum Physics, Key Laboratory of Advanced Optoelectronic Quantum Architecture and Measurement (MOE), School of Physics, Beijing Institute of Technology, Beijing, 100081, China}

	\begin{abstract}
We demonstrate that a class of stable $\mathbb{Z}_2$ monopole charge Dirac point ($\mathbb{Z}_2$DP) phases can robustly exist in real materials, which surmounts the understanding: that is, a $\mathbb{Z}_2$DP is unstable and generally considered to be only the critical point of a $\mathbb{Z}_2$ nodal line ($\mathbb{Z}_2$NL) characterized by a $\mathbb{Z}_2$ monopole charge (the second Stiefel-Whitney number $w_2$)  with space-time inversion symmetry but no spin-orbital coupling. For the first time, we explicitly reveal the higher-order bulk-boundary correspondence in the stable $\mathbb{Z}_2$DP phase. We  propose the alternating-twisted multilayer graphene, which can be regarded as 3D twisted bilayer graphene (TBG), as the first example to realize such stable $\mathbb{Z}_2$DP phase and show that the Dirac points in the 3D TBG are essential degenerate at high symmetric points protected by crystal symmetries and carry a nontrivial $\mathbb{Z}_2$ monopole charge ($w_2=1$), which results in higher-order hinge states along the entire Brillouin zone of the $k_z$ direction. By breaking some crystal symmetries or tailoring interlayer coupling we are able to access $\mathbb{Z}_2$NL phases or other $\mathbb{Z}_2$DP phases with hinge states of adjustable length. In addition, we present other 3D materials which host $\mathbb{Z}_2$DPs in the electronic band structures and phonon spectra. We construct a minimal eight-band tight-binding lattice model that captures these nontrivial topological characters and furthermore tabulate all possible space groups to allow the existence of the stable $\mathbb{Z}_2$DP phases, which will provide direct and strong guidance for the realization of the $\mathbb{Z}_2$ monopole semimetal phases in electronic materials, metamaterials and electrical circuits, etc.
	\end{abstract}
 
	\maketitle
	
	\paragraph{\textcolor{blue}{Introduction.}\textemdash{}}
The breakthrough in magic angle twisted bilayer graphene (TBG) makes it clear that the twist, as a powerful control method, can dramatically manipulate the physical properties of layered materials \cite{caoNature2018a, caoNature2018b, yankowitzScience2019a, serlinScience2020}. In the rapid development of this field, numerous new twisted systems have been experimentally prepared, such as twisted trilayer graphene \cite{parkTunableStronglyCoupled2021, haoElectricField2021}, twisted double-bilayer graphene \cite{shenCorrelatedStatesTwisted2020, liuTunable2020}, alternating- twisted four-layer and five-layer graphene \cite{parkRobust2022a}, twisted transition metal dichalcogenide \cite{wangCorrelatedElectronicPhases2020}, twisted hexagonal boron nitride \cite{niSoliton2019, suTuning2022}, etc. It opens exciting possibilities for engineering exotic quantum states by the twist. A variety of novel quantum states are predicted or observed experimentally in the twisted systems, including unconventional superconducting states \cite{caoNature2018b, yankowitzScience2019a}, topological superconducting states \cite{liuChiralSpinDensity2018, chewHigherOrderTopological2021}, quantum anomalous Hall states \cite{serlinScience2020}, quantum spin Hall states \cite{wuTopologicalInsulators2019}, high order topological insulating states \cite{parkPRL2019b,liuHigherOrderBandTopology2021a}, and so on \cite{xieFractionalChernInsulators2021,ledwithTBNotTB2021,zhaiLayer2022a,wuThreedimensional2020,LedeXNL2021}.

	\begin{figure}
	\begin{center}
		\includegraphics[width=1\linewidth]{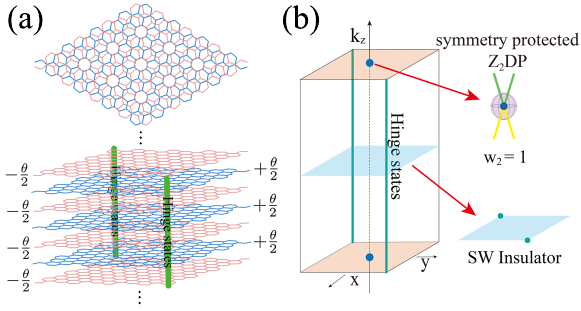}
	\end{center}
	\caption{(a) Top view (top panel) of an alternating-twisted multilayer graphene and its front view (bottom panel) with hinge states. The twisted angles of adjacent layers have same magnitude but opposite direction. (b) Schematic of the stable $\mathbb{Z}_2$ monopole Dirac points protected by crystalline symmetry and the corresponding higher-order topology. The stable Dirac points are at the high symmetric points (blue dots) and carry a nontrivial $\mathbb{Z}_2$ monopole charge. Each 2D $k_z$ plane except $k_z = \pm\pi$ can be viewed as a 2D Stiefel-Whitney insulator, which has two zero modes (green dots) at a pair of $\mathcal{P}\mathcal{T}$-related corners. These corner zero modes make up the hinge states (green lines).}  \label{fig:1}
	\end{figure}

Topological semimetals (TSMs)\cite{bernevigRecentProgressStudy2018, lvExperimental2021, fangTopological2016, linTPRB2018a, wiederNC2020a, fangPRB2021b,chenField2021,ghorashiHigher2020b,qiuHigherOrder2021, wangHigherorder2022, zengTopological2022,zhangZ2Dirac2022,ezawaHigherOrder2018} are materials whose band structures own gap-closing points, lines, or surfaces near the Fermi level. Recent studies show that the TSMs with nodal lines \cite{fangPRB2015, ahnPRL2018a, zhaoPRL2017a, wangPRL2019a, chenPRL2022a}, or Dirac points (DPs) \cite{zhaoPRL2017a, wangPRL2020, chenPRL2022a} can bear a 2D topological invariant called $\mathbb{Z}_2$ monopole charge protected by the space-time inversion ($\mathcal{P}\mathcal{T}$) symmetry in the absence of spin-orbital coupling. The topology of the $\mathbb{Z}_2$ monopole charge Dirac semimetals ($\mathbb{Z}_2$DSMs) is characterized by the second Stiefel-Whitney (SW) number $w_2$ (also called as the real Chern number) \cite{ShiozakiPRB2017, ahnPRL2018a, ahnCPB2019a, zhaoPRL2017a}. Unlike the conventional Dirac semimetals, which do not belong to any of the four common characteristic classes, i.e., Chern class, Stiefel-Whitney class, Pontrjagin class, and Euler class, the $\mathbb{Z}_2$DSMs belong to the Stiefel-Whitney class \cite{nakahara2003geometry, ahnPRL2018a}. Previous studies mainly focused on the $\mathbb{Z}_2$ monopole charge nodal line ($\mathbb{Z}_2$NL) semimetals \cite{ahnPRL2018a, wangPRL2020, chenPRL2022a}, since the $\mathbb{Z}_2$NLs are doubly charged, characterized by 1D winding number and the second Stiefel-Whitney number, and the two topological charges result in different boundary states at distinct boundaries, i.e., 2D drumhead surface states and 1D hinge states. In contrast, the $\mathbb{Z}_2$ monopole charge Dirac point ($\mathbb{Z}_2$DP) phase was considered a critical phase in the evolution of $\mathbb{Z}_2$NLs, unstable in real materials, and having only surface Fermi arcs. However, such surface Fermi arcs are not topologically protected \cite{kargarianPNAS2016}.

In this Letter, we demonstrate the stability of $\mathbb{Z}_2$DPs with crystal symmetries and clearly show the topological-protected robust hallmark higher-order bulk-boundary correspondence in the $\mathbb{Z}_2$DP phase. We predict the alternating-twisted multilayer graphene (ATMG), which is plotted in Fig. \ref{fig:1}(a) and considered as 3D TBG, as the first example of such stable $\mathbb{Z}_2$DSM materials from density functional theory (DFT) calculations and analytic analysis. We take the  ATMG with a large twist angle (21.78$^{\circ}$) and thus strong intervalley scattering as an instance to explicitly show the DPs, $\mathbb{Z}_2$ monopole charge, and higher-order hinge states. The stable DPs are protected by $\mathcal{P}\mathcal{T}$ and other crystalline symmetry operations. We build the effective models for the 3D TBGs. 
By applying strain or pressure we are able to access $\mathbb{Z}_2$NLs or introduce another pair of $\mathbb{Z}_2$DPs resulting in the hinge states with adjustable length. Furthermore, we generalize our discussion, tabulate all possible space groups supporting the stable $\mathbb{Z}_2$DPs and present the corresponding effective models. We suggest such stable $\mathbb{Z}_2$DSMs can also be realized in phonon and metamaterials, such as acoustics, photonics, and electrical circuits, with the allowable space groups.

	\paragraph{\textcolor{blue}{Geometry and Symmetry.}\textemdash{}} We first introduce the crystal structures and symmetry of 2D TBGs. The TBG is constructed by rotating the two layers of $AA$-stacked bilayer graphene around the center of the hexagonal lattice by $-\theta /2$ and $+\theta /2$, respectively. For generic $\theta$, the translation symmetry is broken by the twist. The moir\'e translational symmetry is retained for the specific twist angles, which can take the form of $\theta(m, n)=\operatorname{arccos}[ (3 m^{2}+3 mn+n^{2} / 2)/(3 m^{2}+3 mn+n^{2})]$,
where $m$ and $n$ are coprime positive integers \cite{SantosPRB2012}. The corresponding lattice constant of the moir\'e unit cell is $L=a \sqrt{\left(3 m^{2}+3 m n+n^{2}\right) /[\operatorname{gcd}(n, 3)]}$, where $a$ is the original lattice constant and gcd represents the greatest common divisor. Then, we consider a structure of ATMG where the twisted angles of adjacent layers have the same magnitude but opposite direction as shown in Fig. \ref{fig:1}(a). The ATMG can be viewed as a 3D TBG with two layers of graphene in each unit cell.

	The 2D TBG crystalizes in the hexagonal symmorphic space group P622 with $C_{6z}$ and $C_{2x}$ symmetry about the out-of-plane $z$ and in-plane $x$ axes but no inversion symmetry ($\mathcal{P}$). The 3D TBG belongs to the nonsymmorphic space group P6/mmc (No. 192), which includes $C_{6z}$, $\mathcal{P}$ and $C_{2xy}^{'}=\lbrace C_{2xy} \vert 0 0 \frac{1}{2} \rbrace$. Stacking gives 3D TBG some symmetry operations that 2D TBG does not have, which dramatically affects the topology and band degeneracy of the system.

	\begin{figure}
		\begin{center}
			\includegraphics[width=1\linewidth]{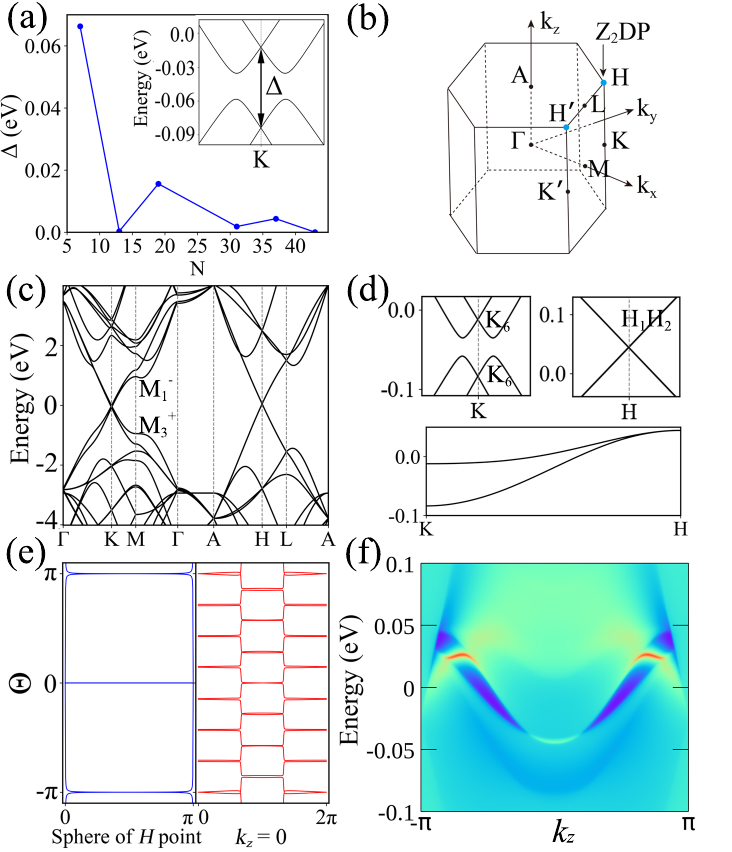} 
		\end{center} 
		\caption{(a) The band gap at $K$ point versus the commensuration cells size $N$. The inset shows the band structure near $K$ point of the 3D TBG with $N = 7$ ($\theta = 21.78^{\circ}$) (b) Brillouin zone and the high symmetric points. Blue dots represent the Dirac points at $H/H'$ points.  (c) Band structure of the 3D TBG with $\theta = 21.78^{\circ}$ from DFT calculation. M$_3^+$ and M$_1^-$ are representations of the valence and conduction bands at $M$. (d) The magnified view of the regions near $K$, $H$, and high symmetric line $KH$. K$_6$ and H$_1$H$_2$ are band representations at $K$ and $H$. (e) Wilson loop spectrum for the 3D TBG on the sphere enclosing H point (blue lines) and torus of $k_z=0$ (red lines), respectively. (f) Hinge Fermi arcs of the 3D TBG along the $k_z$ direction.} \label{fig:2}
	\end{figure}

	\paragraph{\textcolor{blue}{Band structure, $\mathbb{Z}_2$ topology and higher-order bulk-boundary correspondence.}\textemdash{}} In stark contrast to the small twist angle limit ($\lesssim1^{\circ}$),  the $U(1)$ valley symmetry in TBGs is broken at a large angle with a gap opened at $K$ point due to the intervalley scattering\cite{shallcrossPRL2008,parkPRL2019b}. The size of the gap at $K$ depends on the size of the commensuration cell $N$ with $N = (L/a)^2$ and decays rapidly as $N$ increases, as shown in Fig. \ref{fig:2}(a), and the 3D TBG with $N=7$ has the largest band gap at $K$ point, about 60 meV. Figure \ref{fig:2}(c) shows the DFT band structure of $\sqrt{7} \times \sqrt{7}$ ($N=7$, $\theta = 21.78^{\circ}$) 3D TBG. The magnified views of the regions near $K$, $H$, and high symmetric line $KH$ are plotted in Fig. \ref{fig:2}(d). The corresponding band representations are also given. The band gap is about 20 meV near $K$. As the $k_z$ increases, the band gap becomes smaller and smaller and finally the bands close at $H$. The bands are double degenerate along the $KH$ line protected by the $C_{3z}$ and $\mathcal{PT}$ symmetry and become a four-fold degenerate point at $H$. The band gap of 3D TBG around $K$ point is dramatically affected by the layer distance. Under pressure, the band gap near $K$ can reach 0.1 eV at the layer distance of 2.95 \AA\ (3.3 \AA\ without pressure) [See details in Supplemental Material (SM) \cite{SM}]. 
	
	The $\mathbb{Z}_2$ monopole topology for a $\mathbb{Z}_2$DP or $\mathbb{Z}_2$NL can be characterized by the second Stiefel-Whitney number $w_2$, which can be calculated efficiently by using the Wilson loop method \cite{ahnPRL2018a}. We calculate the Wilson loop of the sphere enclosing the DP ($H$ point), as shown in the left panel of Fig. \ref{fig:2}(e). This Wilson loop spectrum exhibits $w_2 = 1$ with the characteristic winding of a $\mathbb{Z}_2$DP, as it only has one crossing point on $\Theta = \pi$. Normally, a nontrivial $\mathbb{Z}_2$NL can shrink to a $\mathbb{Z}_2$DP with only critical parameters. Such $\mathbb{Z}_2$DPs are not stable under the protection of $\mathcal{P}\mathcal{T}$ symmetry.  We point out that one new kind of $\mathbb{Z}_2$DPs can stably exist at certain high symmetric points with additional crystalline symmetry operations forming essential degenerate points, such as in the 3D TBG. These $\mathbb{Z}_2$DPs are even more stable than $\mathbb{Z}_2$NLs because they are pinned at high symmetric points and therefore cannot be annihilated without symmetry broken.

	The ATMGs (3D TBGs) with nontrivial $\mathbb{Z}_2$ monopole topology have a higher-order bulk-boundary correspondence with a hallmark hinge state, which is shown in Fig. \ref{fig:2}(f) and calculated by the recursive hinge Green function method, as described in SM \cite{SM}. To better understand the higher-order bulk-boundary correspondence, we further calculate the Wilson loop spectrum at the planes of $k_z \in(-\pi,\pi)$, with the $k_z=0$ plane shown in the right panel of  Fig. \ref{fig:2}(e). The crossing points on $\Theta = 0$ and $\Theta = \pi$ in the Wilson loop are both odd numbers, which indicates the $w_2 = 1$. Each slice with a specific $k_z$ in the Brillouin zone (BZ) is a torus and can be taken as a 2D subsystem. In the 3D TBG, the entire $k_z$ slices except $k_z =\pm \pi$ carry nontrivial $w_2=1$. Therefore, each slice in the region of (-$\pi, \pi$) is a 2D Stiefel-Whitney insulator, which has a pair of topologically protected corner zero modes, as schematically shown in Fig. \ref{fig:1}(b). Such zero modes from all of these nontrivial $k_z$ slices make up the topological protected hinge states on a pair of $\mathcal{P}\mathcal{T}$-related hinges \cite{note1}.

\paragraph{\textcolor{blue}{Symmetry protected essential degenerate $\mathbb{Z}_2$DPs and effective models}\textemdash{}} 
	At $H/H'$ points of 3D TBG, the $\mathbb{Z}_2$DPs are protected by not only $\mathcal{P}\mathcal{T}$ symmetry but also $C_{3z}^{\pm}$, $\widetilde{\mathcal{\sigma}}_{d}=\lbrace \sigma_{d} \vert 0 0 \frac{1}{2} \rbrace$ and $M_z$. We first demonstrate an essential degenerate DP at $H$ with these symmetry operations. The algebra of these symmetry operations can be written as $(M_z\mathcal{P}\mathcal{T})^2\equiv \mathcal{A}^2=1, \widetilde{\mathcal{\sigma}}_{d}^2=-1, C_{3z}^{\pm}\mathcal{A}=\mathcal{A}C_{3z}^{\pm}, C_{3z}^{\pm}\widetilde{\mathcal{\sigma}}_{d}=\widetilde{\mathcal{\sigma}}_{d}C_{3z}^{\mp}, \mathcal{A}\widetilde{\mathcal{\sigma}}_{d}=\widetilde{\mathcal{\sigma}}_{d}\mathcal{A}$ \cite{SM}.
The Bloch states can be chosen as the eigenstates of $C_{3z}^{+}$, denoted as $|\phi\rangle$ with the eigenvalues $\phi=1, e^{\pm i\frac{2\pi}{3}}$. Since $C_{3z}^{+}$ commutes with $\mathcal{A}$ and $\mathcal{A}i = -i$, the two states $|e^{i\frac{2\pi}{3}}\rangle$ and $\mathcal{A}|e^{i\frac{2\pi}{3}}\rangle$ would be degenerate, as $C_{3}^{+}\mathcal{A}|e^{i\frac{2\pi}{3}}\rangle=e^{-i\frac{2\pi}{3}}\mathcal{A}|e^{i\frac{2\pi}{3}}\rangle$. Similarly, the two states $\widetilde{\mathcal{\sigma}}_{d}|e^{i\frac{2\pi}{3}}\rangle$ and $\widetilde{\mathcal{\sigma}}_{d}\mathcal{A}|e^{i\frac{2\pi}{3}}\rangle$ are degenerate. Since $(\widetilde{\mathcal{\sigma}}_{d}\mathcal{A})^2=-1$ and $\langle e^{i\frac{2\pi}{3}}|\widetilde{\mathcal{\sigma}}_{d}\mathcal{A}|e^{i\frac{2\pi}{3}}\rangle$ = 0, the two degenerate states $|e^{i\frac{2\pi}{3}}\rangle$ and $\mathcal{A}|e^{i\frac{2\pi}{3}}\rangle$ and their Kramers-like partner $\widetilde{\mathcal{\sigma}}_{d}\mathcal{A}|e^{i\frac{2\pi}{3}}\rangle$ and $\widetilde{\mathcal{\sigma}}_{d}|e^{i\frac{2\pi}{3}}\rangle$ are linearly independent. Consequently, the four states $\lbrace|e^{i\frac{2\pi}{3}}\rangle$, $\mathcal{A}|e^{i\frac{2\pi}{3}}\rangle$, $\widetilde{\mathcal{\sigma}}_{d}|e^{i\frac{2\pi}{3}}\rangle$, $\widetilde{\mathcal{\sigma}}_{d}\mathcal{A}|e^{i\frac{2\pi}{3}}\rangle\rbrace$ must be degenerate at the same energy, forming an essential degenerate DP.

\begin{figure}
		\begin{center}
			\includegraphics[width=1\linewidth]{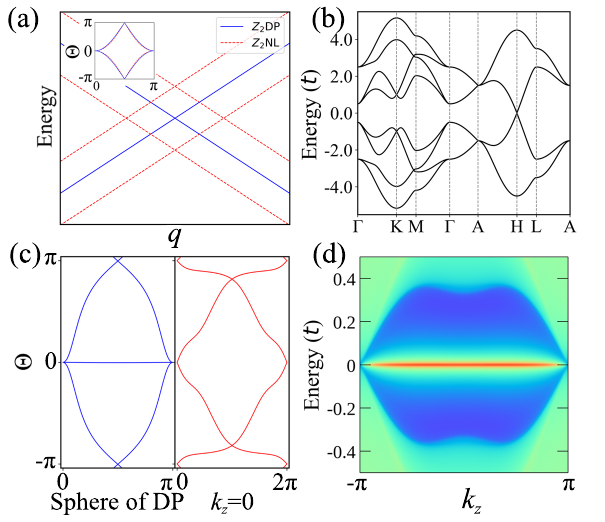}
		\end{center}
		\caption{(a) Band structures of the $\mathbf{k}\cdotp \mathbf{p}$ model without/with a perturbation term (blue solid/red dashed lines), which indicate a Dirac point and nodal line, respectively. The inset shows the respective Wilson loops. (b) The Band structure of the minimal TB lattice model. (c) Wilson loops of a sphere enclosing the H point and a torus ($k_z = 0$ plane), respectively. (d) Hinge states of the TB model in the $Z_2$DP phase.} \label{fig:3}
	\end{figure}

Constrained by these symmetry operations \cite{SM}, the $\mathbf{k}\cdotp \mathbf{p}$ model around $H$ expanded to the first order of $q=k-H$ reads 
\begin{equation}
H_{DP}=\alpha(q_{x}\Gamma_{x,z}-q_{y}\Gamma_{y,0})+q_{z}(\beta_{1}\Gamma_{x,x}+\beta_{2}\Gamma_{y,x}),
\end{equation}
where $\alpha$ and $\beta_{i}$ are real parameters and $\Gamma_{i, j} = \sigma_i \otimes \sigma_j$. The energy eigenvalues are $ E_{DP}=\pm\sqrt{\alpha^{2}\rho^{2}+\beta^{2}q_{z}^{2}\pm2\alpha\left|\beta_{2}q_{z}\right|\rho}$ with $\rho=\sqrt{q_x^2+q_y^2}$ and $\beta=\sqrt{\beta_{1}^2+\beta_{2}^2}$. One can see the four-fold degenerate DP located at $q_x=q_y=q_z=0$ [Fig. \ref{fig:3}(a)]. To confirm the $\mathbb{Z}_2$ topological charge of the model, we calculate the Wilson loop of a sphere enclosing the DP, which is nontrivial with $w_2=1$ [Fig. \ref{fig:3}(a)]. A perturbation term $m_0 \sigma_{0} \otimes\sigma_{z}$, which breaks the $\widetilde{\mathcal{\sigma}}_{d}$, is added on the $H_{DP}$ and the energy eigenvalues are $E_{NL}=\pm\sqrt{\left(\sqrt{\beta_{2}^{2}q_{z}^{2}+m_{0}^{2}}\pm\alpha\rho\right)^{2}+\beta_{1}^{2}q_{z}^{2}}$. One can see that the valence and conduction bands touch at $q_z = 0$ and $\rho = \lvert m_0/\alpha\rvert$, indicating that the $\mathbb{Z}_2$DP is split into a NL [Fig. \ref{fig:3}(a)]. Moreover, the $\mathbb{Z}_2$ monopole charge is preserved in the NL, resulting in a  $\mathbb{Z}_2$NL [Fig. \ref{fig:3}(a)]. The other NLs ($\rho=0$) from two valence or conduction bands link with the $\mathbb{Z}_2$NL.

    To further explore the higher-order bulk-boundary correspondence of the $\mathbb{Z}_2$DPs and get a better fitting with the 3D TBG in the band representation, we construct a minimal tight-binding (TB) lattice model. The model assumes $d_{xz}$ and $d_{yz}$ symmetry orbitals at the Wyckoff position 4$d$ of a hexagonal lattice with nonsymmorphic space group P6/mmc. This model can be viewed as two layers of honeycomb lattice in a unit cell. The intra-layer hopping integrals between $d_{xz, yz}$-like orbitals on each layer of the honeycomb lattice are constructed via the Slater-Koster formalism, which reflects coexisting $\sigma$ and $\pi$ bonds. 
    
    The intra-layer Hamiltonian with only nearest neighbor hopping of each layer reads
    \begin{equation}
H_{intra}=\sum_{i \mu, j \nu} t_{i \mu, j \nu} c_{i \mu}^{\dagger} c_{j \nu},
    \end{equation}    
    where $\mu, \nu = x, y$ represent the $d_{xz}$ and $d_{yz}$ orbitals, $i$, $j$ stand for the two sublattices of one layer honeycomb lattice. The hopping integrals $t_{i \mu, j \nu}$ read
    \begin{equation}
t_{i \mu, j \nu}=t_\sigma^{i j} \cos \theta_{\mu, i j} \cos \theta_{\nu, i j}+t_\pi^{i j} \sin \theta_{\mu, i j} \sin \theta_{\nu, i j},
    \end{equation}
    where $\theta_{\mu, i j}$ represents the angle between the direction of $\mu$ and $\mathbf{r}_j-\mathbf{r}_i$ \cite{SM}. The Slater-Koster parameters $t_{\sigma / \pi}^{i j}$ denote the hopping integrals contributed by $\sigma / \pi$ bonds. The interlayer hopping has the form of
    \begin{equation}
    H_{inter} = r_2 \cos \left(\frac{k_z}{2}\right) s_{0} \sigma_{0} \tau_{x} + r_1 \cos \left(\frac{k_z}{2}\right) s_{y} \sigma_{0} \tau_{y}
    \end{equation}
	where the Pauli matrices $s$, $\sigma$, and $\tau$ act on the orbital, sublattice, and layer degree of freedom, respectively. The $r_1$ and $r_2$ denote the hopping integrals between the orbitals in different layers. 
	
	Therefore, the minimal eight-band model reads
	\begin{equation}
	H_{8} = H_{intra} \tau_{0} + H_{inter}.
	\end{equation}
	The Hamiltonian belongs to the space group P6/mmc, which is demonstrated in SM \cite{SM}. The band structure shows that a couple of DPs are pinned at $H/H'$ and at the Fermi level [Fig. \ref{fig:3}(b)]. The degeneracy at $H$ and the band representations are both consistent with the 3D TBG.
The nontrivial monopole charge of the DPs are confirmed by the Wilson loop [left panel of Fig. \ref{fig:3}(c)]. Similar to the above analysis, each slice in the region of (-$\pi, \pi$) is a 2D Stiefel-Whitney insulator [right panel of Fig. \ref{fig:3}(c)], whose corner zero modes constitute the hinge Fermi arc, as shown in Fig. \ref{fig:3}(d).

We also construct a Slater-Koster TB model with only $p_z$ orbitals of carbon, which has good agreements with the DFT results \cite{SM}.

\begin{figure}
		\begin{center}
			\includegraphics[width=1\linewidth]{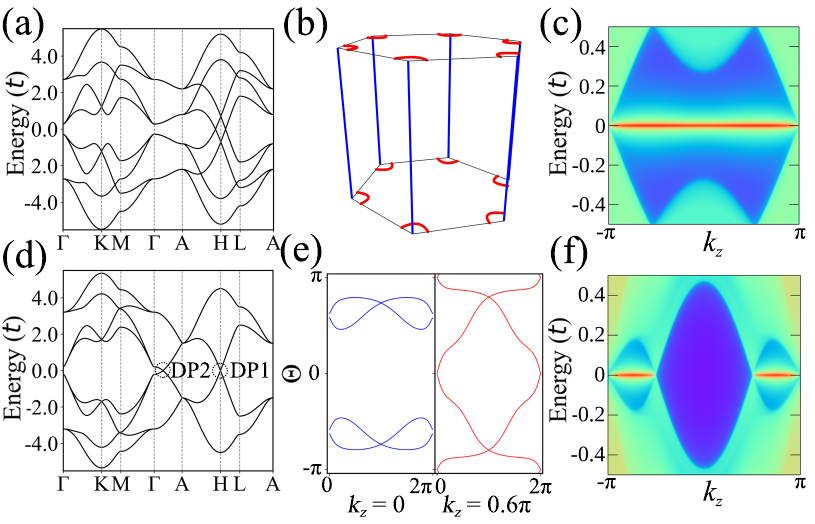}
		\end{center}
		\caption{(a) Band structure of the TB lattice model in a $\mathbb{Z}_2$NL phase. (b) Distribution of the NLs with a special linking structure. The red circles are the $\mathbb{Z}_2$NLs and the blue lines are NLs from the two highest valence bands or two lowest conduction bands. (c) Hinge states of the TB model in the $\mathbb{Z}_2$NL phase. (d) Band structure of the minimal TB model with two pairs of $\mathbb{Z}_2$DPs. (e) Wilson loops of two typical $k_z$ planes. (f) Hinge states of the TB model with two pairs of $\mathbb{Z}_2$DPs.} \label{fig:4}
	\end{figure}

{\textcolor{blue}{Manipulation of $\mathbb{Z}_2$ topological quantum states}\textemdash{}} 
One can induce novel topological child phases from the $\mathbb{Z}_2$DP parent phase. Adding different onsite energy of the two layers in the minimal TB model, the symmetry $C_{2xy}^{'}$ is broken and a $\mathbb{Z}_2$NL emerges and links with other NLs formed by two valence or conduction bands, as shown in Figs. \ref{fig:4}(a, b) \cite{SM}. Similar to the DP phase case, each slice of $k_z \in(-\pi,\pi)$ is a 2D Stiefel-Whitney insulator with a pair of topologically protected corner zero modes. These zero modes will constitute the topologically protected hinge states, as shown in Fig. \ref{fig:4}(c). Such scenario to induce the $\mathbb{Z}_2$NL phase can be realized in 3D TBG with uniaxial strain applied, as demonstrated in SM \cite{SM}.

Tailoring the parameters of interlayer coupling can result in another pair of DPs along the high symmetric line $\Gamma A$ in addition to the pair of DPs at the points $H/H'$, which are labeled as DP2 and DP1 respectively [Fig. \ref{fig:4}(d)]. The DP2 is an accidental degenerate point while the DP1 is an essential degenerate point. Both types of DPs have nontrivial $\mathbb{Z}_2$ topology. The $w_2$ of the $k_z$ slices between two accidental degenerate $\mathbb{Z}_2$DPs become trivial [Fig. \ref{fig:4}(e)], and the hinge states are split into two pieces [Fig. \ref{fig:4}(f)] \cite{SM}.  As a result, one can tailor the length of the hinge Fermi arc by tuning the interlayer hopping parameters.

\begin{table}[h]
  \centering
  {\tabcolsep0.05in
  \caption{List of all possible space groups with essential or accidental degenerate $\mathbb{Z}_2$ monopole Dirac points ($\mathbb{Z}_2$DPs) and the corresponding momentum distribution.}
    \begin{tabular}{cc}
    \hline \hline
          & SG Number \\ \hline
    \multirow{2}[0]{*}{Essential  $\mathbb{Z}_2$DPs} & 73 (W),  142 (P), 165(H), \\
                                             & 192(H), 206(P), 230(P) \\
    Accidental  $\mathbb{Z}_2$DPs & 175 ($\Gamma A$), 176 ($\Gamma A$), 191-194 ($\Gamma A$) \\ \hline \hline
    \end{tabular}
  \label{table1}
  }
\end{table}	
	
\paragraph{\textcolor{blue}{Stable $\mathbb{Z}_2$DPs in all possible space groups beyond 3D TBG.}\textemdash{}} Since the DPs are protected by the crystal symmetry at (along) high-symmetric points (lines), we can check their topology of $\mathbb{Z}_2$ monopole charge by calculating the Wilson loop of all DPs in the 230 type-II magnetic space groups \cite{bradley2010mathematical,TangPRB2021,yuScibull2021}. Finally, we find six space groups can protect essential degenerate $\mathbb{Z}_2$DPs and six space groups can protect accidental degenerate $\mathbb{Z}_2$DPs, as given in Table \ref{table1}. The corresponding $\mathbf{k}\cdot\mathbf{p}$ effective models, band structures, and Wilson loop spectra are presented in SM \cite{SM}.

The $\mathbb{Z}_2$DPs are widely present in the electronic band structures and phonon spectra of materials which belong to these space groups. For example, besides the ATMG, the $\mathbb{Z}_2$DPs are also present in the band structure at the $P$ point of Si~\cite{Si206} with the space group No. 206, the $H$ point of the phonon spectrum in LaF$_3$~\cite{LaF3165}  with the space group No. 165, and in the phonon spectrum of KSn~\cite{KSn142} with the space group No. 142, as shown in SM \cite{SM}. Moreover, one can also construct metamaterials such as photonic and phononic crystals to realize $\mathbb{Z}_2$DP phases based on these space groups.

	\paragraph{\textcolor{blue}{Discussion.}\textemdash{}} We demonstrate that $\mathbb{Z}_2$DPs can stably exist in real materials and give all possible space groups to allow the existence of $\mathbb{Z}_2$DPs. The nontrivial $\mathbb{Z}_2$ monopole charge topology is characterized by the second Stiefel-Whitney number $w_2$. Our research shows that the $\mathbb{Z}_2$DP phase is stable and even can be observed more readily in experiments than the $\mathbb{Z}_2$NL phase. This is because the $\mathbb{Z}_2$NL phase easily undergoes the pair annihilation, whereas the $\mathbb{Z}_2$DPs stably exist in specific high-symmetric points for all materials in the twelve allowable space groups, which we point out explicitly. Specifically, we propose ATMGs as the first example of such stable $\mathbb{Z}_2$DSM with higher-order hinge Fermi arcs, which can be probed by scanning tunneling spectroscopy, as exploring the higher-order topology in Bismuth \cite{schindlerBismuth2018}. The $\mathbb{Z}_2$DSM in 3D TBG enriches the topological phases in twistronics. Based on our effective models and proposed list of allowed space groups, the new and stable kinds of $\mathbb{Z}_2$DSM phases are expected to be realized in metamaterials, such as acoustics, photonics, and electrical circuits, thanks to the flexibility of the building blocks.

\paragraph{\textcolor{blue}{Acknowledgments.}\textemdash{}} 
		The work is supported by the National Key R\&D Program of China (Grant No. 2020YFA0308800) and the NSF of China (Grants No. 11922401).

	\bibliographystyle{apsrev4-2}
	\bibliography{reference}

\begin{thebibliography}{66}
\bibitem [{not()}]{note1}%
  \BibitemOpen
\bibfield  {journal} {  }\href@noop {} {\bibinfo  {journal} {Note that our
  calculation based on the recursive hinge Green function and Wannier function
  gives one hinge state [Fig. \ref{fig:2}(f)]. A pair of hinge states are
  obtained by direct diagonalization of a tubelike sample \cite{SM}}\
  }\BibitemShut {NoStop}%
\end{thebibliography}

\end{document}


\preprint{APS/123-QED}

\title{Supplemental Material \textemdash{} Stable Higher-Order Topological Dirac Semimetals with $\mathbb{Z}_2$ Monopole Charge in Alternating-twisted Multilayer Graphenes and beyond}
\author{Shifeng Qian}
\affiliation{Key laboratory of advanced optoelectronic quantum architecture and measurement (MOE), School of Physics, Beijing Institute of Technology, Beijing 100081, China }

\author{Yongpan Li}
\affiliation{Key laboratory of advanced optoelectronic quantum architecture and measurement (MOE), School of Physics, Beijing Institute of Technology, Beijing 100081, China }

\author{Cheng-Cheng Liu}
\email{ccliu@bit.edu.cn}
\affiliation{Key laboratory of advanced optoelectronic quantum architecture and measurement (MOE), School of Physics, Beijing Institute of Technology, Beijing 100081, China }


\maketitle

\tableofcontents

	\section{Computational methods}
The calculations of the structure optimization and band structures of 3D twisted bilayer graphene (TBG) are performed using DFT in the Perdew-Becke-Ernzerhof (PBE) generalized gradient approximation (GGA) \cite{PBE} implemented in the Vienna $ab$ $initio$ simulation package (VASP) \cite{VASP}.  The Brillouin zone is sampled by a 12 $\times$ 12 $\times$ 12 mesh and the plane-wave energy cutoff is 600 eV. In the structure optimization, van der Waals interactions between the layers are considered by the DFT-D3 method \cite{DFTD3}. The symmetry-dapted Wannier functions are constructed using the WANNIER90 code \cite{wannier1} and VASP. Based on the constructed Wannier functions, we obtain the hinge states and Wilson loop.  The representations of electronic states from DFT results are calculated using irvsp code \cite{irvsp}.

    \section{Effective model of $\mathbb{Z}_2$DP at the $H/H'$ points of Space Group No. 192}
    \label{Section_Heff_192}		
     At $H$/$H'$ points of 3D TBG, the $\mathbb{Z}_2$ Dirac points ($\mathbb{Z}_2$DPs) are protected by $C_{3z}$ ($(x, y, z) \rightarrow (-x+y, -x, z)$), $\widetilde{\mathcal{\sigma}}_{d}=\lbrace \sigma_{d} \vert 0 0 \frac{1}{2} \rbrace$ ($(x, y, z) \rightarrow (y, x, z+1/2)$), $M_z$ ($(x, y, z) \rightarrow (x, y, -z)$) and $\mathcal{P}\mathcal{T}$ symmetries. The algebra of these symmetries can be written as 
\begin{equation}
\begin{split}
&(\mathcal{P}\mathcal{T})^2 = 1, \widetilde{\mathcal{\sigma}}_{d}^2=-1, M_{z}^2=1,\\
&C_{3z}M_z=M_zC_{3z},  C_{3z}(\mathcal{P}\mathcal{T})=(\mathcal{P}\mathcal{T})C_{3z},  C_{3z}^{\pm}\widetilde{\mathcal{\sigma}}_{d}=\widetilde{\mathcal{\sigma}}_{d}C_{3z}^{\mp}, \\
&M_z\widetilde{\mathcal{\sigma}}_{d}=-\widetilde{\mathcal{\sigma}}_{d}M_z, M_z(\mathcal{P}\mathcal{T}) = (\mathcal{P}\mathcal{T})M_z, (\mathcal{P}\mathcal{T})\widetilde{\mathcal{\sigma}}_{d}=-\widetilde{\mathcal{\sigma}}_{d}(\mathcal{P}\mathcal{T}).
\end{split}
\end{equation}
According to these algebra of these symmetries, the basis can be chosen as \{$\left|e^{-i\frac{2\pi}{3}},1\right\rangle$, $\widetilde{\mathcal{\sigma}}_{d}\left|e^{-i\frac{2\pi}{3}},1\right\rangle ,\mathcal{PT}\left|e^{-i\frac{2\pi}{3}},1\right\rangle$, $\mathcal{PT}\widetilde{\mathcal{\sigma}}_{d}\left|e^{-i\frac{2\pi}{3}},1\right\rangle $\}, where $\left|e^{-i\frac{2\pi}{3}},1\right\rangle$ is the eigenstate of both $C_{3z}$ and $M_z$ with respective eigenvalues $e^{-i\frac{2\pi}{3}}$ and $1$, since the $C_{3z}$ commutes with $M_z$.
     The representations of the above symmetry operators can be obtained by the operators acting on the basis, for example,
    \begin{equation}
	\begin{split}
     C_{3}\left(\widetilde{\mathcal{\sigma}}_{d}\left|e^{-i\frac{2\pi}{3}},1\right\rangle \right)&=\widetilde{\mathcal{\sigma}}_{d}C_{3}^{-1}\left|e^{-i\frac{2\pi}{3}},1\right\rangle =e^{i\frac{2\pi}{3}}\left(\widetilde{\mathcal{\sigma}}_{d}\left|e^{-i\frac{2\pi}{3}},1\right\rangle \right), \\
     M_{z}\left(\widetilde{\mathcal{\sigma}}_{d}\left|e^{-i\frac{2\pi}{3}},1\right\rangle \right)&=-\widetilde{\mathcal{\sigma}}_{d}M_{z}\left|e^{-i\frac{2\pi}{3}},1\right\rangle =-\left(\widetilde{\mathcal{\sigma}}_{d}\left|e^{-i\frac{2\pi}{3}},1\right\rangle \right), \\
     \widetilde{\mathcal{\sigma}}_{d}\left(\widetilde{\mathcal{\sigma}}_{d}\left|e^{-i\frac{2\pi}{3}},1\right\rangle \right)&=-\left|e^{-i\frac{2\pi}{3}},1\right\rangle , \\
     \mathcal{PT}\left(\mathcal{PT}\left|e^{-i\frac{2\pi}{3}},1\right\rangle\right)& =\left|e^{-i\frac{2\pi}{3}},1\right\rangle.
     \end{split}
     \end{equation}
     Therefore, the corepresentations of these symmetry operators can be written as
     \begin{equation}
\begin{split}
D(C_{3z}) = &\frac{-\sigma_{0} \otimes \sigma_{0}-i\sqrt{3}\sigma_{z} \otimes \sigma_{z}}{2} ,\\
D(\widetilde{\mathcal{\sigma}}_{d})=&i\sigma_{z} \otimes \sigma_{y}, \\
D(M_z) = &\sigma_{0} \otimes \sigma_{z},\\
 D(\mathcal{P}\mathcal{T})=&\sigma_{x} \otimes \sigma_{0}K,
\end{split}
     \end{equation}
where $K$ is the complex conjugation.
The Hamiltonian satisfies the equation of symmetry constraint
\begin{equation}
      D(R)^{-1}H(k)D(R)=H(R^{-1}k), 
\end{equation}
  where $R$ represents the above four symmetry operators.
     Finally, constrained by these symmetries, the effective
Hamiltonian around the $H$ point expanded to the first order of $k=H+q$ is 
\begin{equation}
H_{DP}= q_x(\alpha_1 \Gamma_{y, z} +\alpha_2\Gamma_{x, z})+q_y(\alpha_1 \Gamma_{x, 0} -\alpha_2\Gamma_{y, 0})+q_z(\alpha_3 \Gamma_{x, x} +\alpha_4 \Gamma_{y, x}),
\end{equation}
where $\alpha_{i}$ are real parameters and $\Gamma_{i, j} = \sigma_i \otimes \sigma_j$. 
The effective Hamiltonian can be simplified by a unitary transformation $H_{DP}\rightarrow U_{\theta}H_{DP}U_{\theta}^{-1}$ with $U_{\theta}=e^{i\frac{\theta}{2}\Gamma_{z,0}}$. The simplified effective Hamiltonian reads
\begin{equation}\label{Hdp}
H_{DP}=\alpha(q_{x}\Gamma_{x,z}-q_{y}\Gamma_{y,0})+q_{z}(\beta_{1}\Gamma_{x,x}+\beta_{2}\Gamma_{y,x}),
\end{equation}
where $\theta=\arctan\frac{\alpha_{1}}{\alpha_{2}},\alpha=\sqrt{\alpha_{1}^{2}+\alpha_{2}^{2}},\beta_{1}=\frac{\alpha_{2}\alpha_{3}+\alpha_{1}\alpha_{4}}{\alpha},\beta_{2}=\frac{\alpha_{2}\alpha_{4}-\alpha_{1}\alpha_{3}}{\alpha}$.

\begin{figure}[h]
		\begin{center}
			\includegraphics[width=0.6\linewidth]{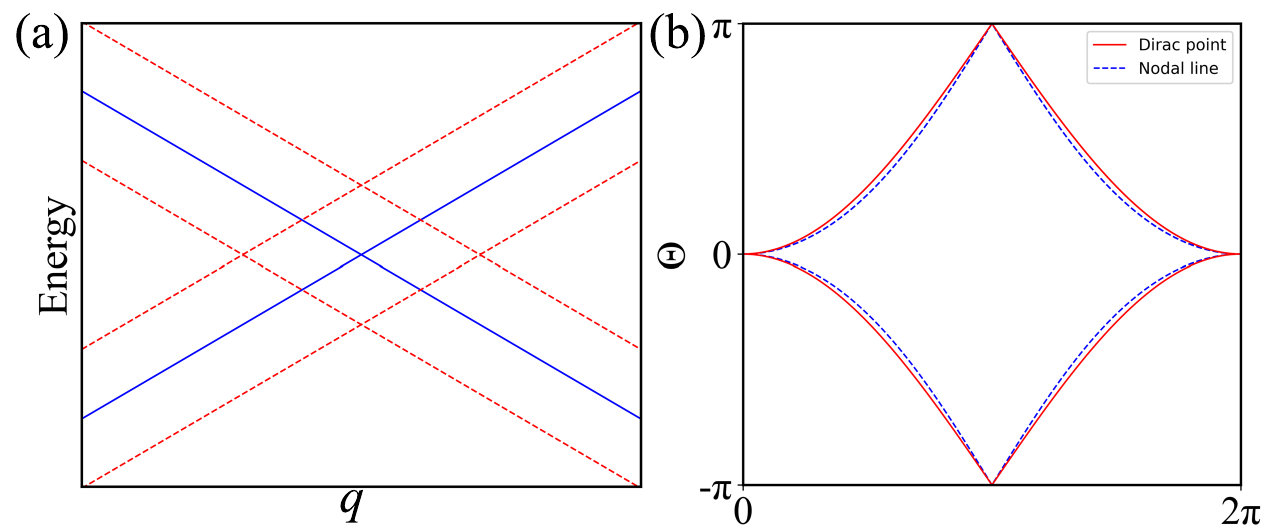} 
		\end{center} 
		\caption{Band structure and Wilson loop of the effective Hamiltonian of the $\mathbb{Z}_2$DP and $\mathbb{Z}_2$NL (a) The band structures of the DP (blue solid line) and the NL  (red dashed line) calculated from effective model $H_{DP}$ and $H_{DP}$ plus a perturbation term. (b) Their Wilson loops with a sphere enclosing the $\mathbb{Z}_2$DP and $\mathbb{Z}_2$NL.} \label{fig:kp}
	\end{figure}

The energy eigenvalues are
\begin{equation}
 E_{DP}=\pm\sqrt{\alpha^{2}\rho^{2}+\beta^{2}q_{z}^{2}\pm2\alpha\left|\beta_{2}q_{z}\right|\rho}, 
 \end{equation}
 where $\rho=\sqrt{q_x^2+q_y^2}$ and $\beta=\sqrt{\beta_{1}^2+\beta_{2}^2}$. One can see a four-fold degenerate DP located at $q_x=q_y=q_z=0$ [Fig. \ref{fig:kp}(a)]. To confirm the $\mathbb{Z}_2$ topological charge of the model, we calculate the Wilson loop of a sphere enclosing the DP [Fig. \ref{fig:kp}(b)]. The crossing point on $\Theta = \pi$ in the Wilson loop is an odd number, which indicates the nontrivial $w_2 = 1$. Since the $\mathbb{Z}_2$ topological charge cannot be annihilated on its own, the evolution of $\mathbb{Z}_2$DP itself will not change its topology. Therefore, the nontrivial topology of the model is not dependent on the parameters.

The $\mathbb{Z}_2$DP can evolve into a $\mathbb{Z}_2$ nodal line ($\mathbb{Z}_2$NL) with some symmetries broken but the $\mathcal{P}\mathcal{T}$. A perturbation term $m_0 \sigma_{0} \otimes\sigma_{z}$, which breaks the $\widetilde{\sigma}_{d}$, is added on the $H_{DP}$ (Eq. \ref{Hdp}) and the energy eigenvalues are 
 \begin{equation}
E_{NL}=\pm\sqrt{\left(\sqrt{\beta_{2}^{2}q_{z}^{2}+m_{0}^{2}}\pm\alpha\rho\right)^{2}+\beta_{1}^{2}q_{z}^{2}}.
 \end{equation}
 One can see that the valence and conduction bands touch at $q_z = 0$ and $\rho = \lvert m_0/ \alpha\rvert$, indicating the $\mathbb{Z}_2$DP is spilled into a $\mathbb{Z}_2$NL [Fig. \ref{fig:kp}(a)]. On the other hand, the two occupied bands or two unoccupied bands are also touching along another NL ($q_x = q_y = 0$) when $\rho = 0$. The $\mathbb{Z}_2$NL links with such NL. Therefore, the $\mathbb{Z}_2$NL is stable and distinct from common NLs because of the special linking structure. The nontrivial $\mathbb{Z}_2$ topology of the NL is also confirmed by the Wilson loop [Fig. \ref{fig:kp}(b)].

\section{Minimal eight-band tight-binding lattice model}
	
\begin{figure}[t]
		\begin{center}
			\includegraphics[width=0.9\linewidth]{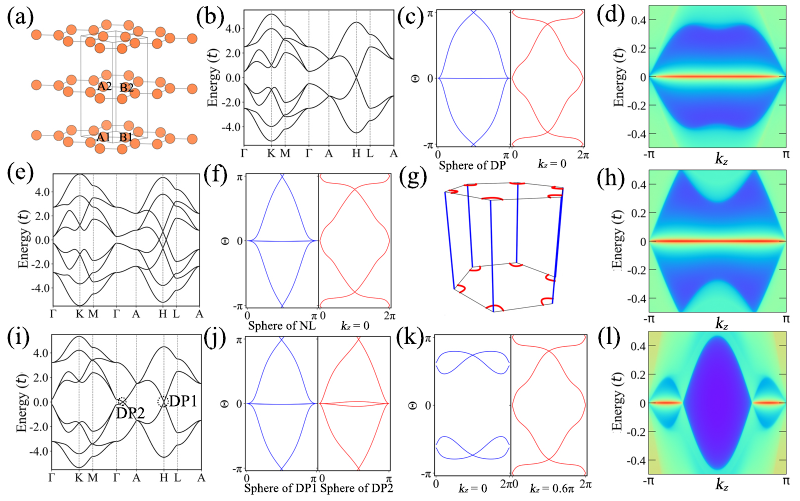} 
		\end{center} 
		\caption{Band structures, Wilson loops, and hinge states of the minimal TB lattice model in different $\mathbb{Z}_2$ topological phases. (a) Lattice and Wyckoff positions 4$d$. The rhombus represents the primitive cell and orange balls represent the Wyckoff positions 4$d$. A1, A2, B1, and B2 denote the two sublattices of two layers. (b) Band structure of the TB model in a $\mathbb{Z}_2$DP phase. (c) Wilson loops of a sphere enclosing the $H$ point and a torus ($k_z=0$ plane), respectively. (d) Hinge states of the TB model in the $\mathbb{Z}_2$DP phase. (e) Band structure of the TB model in a $\mathbb{Z}_2$NL phase. (f) Wilson loops of a sphere enclosing the nodal line near the $H$ point and a torus ($k_z=0$ plane), respectively. (g) Momentum distribution of the NLs with a special linking structure. The red circles are $\mathbb{Z}_2$ NLs from the valence band and the conduction band and the blue lines are NLs from the two highest valence bands or two lowest conduction bands. (h) Hinge states of TB model in the $\mathbb{Z}_2$NL phase. (i) Band structure of the minimal TB model with two pairs of $\mathbb{Z}_2$DPs. (j) Wilson loops of a sphere enclosing the DP1 and DP2, respectively. (k) Wilson loops of two typical $k_z$ planes. (l) Hinge states of the TB model with two pairs of $\mathbb{Z}_2$DPs.} \label{fig:Model}
	\end{figure}

To further explore the higher-order bulk-boundary correspondence of the $\mathbb{Z}_2$DPs, we construct a minimal tight-binding (TB) model. To get a better fitting with the 3D TBG both in the shape and representation of the low-energy bands, the model assumes spinless particles which belong to $d_{xz}$ and $d_{yz}$ symmetry on a hexagonal lattice with nonsymmorphic space group P6/mmc at the Wyckoff position 4$d$. This model can be viewed as two layers of honeycomb lattice in a unit cell. The intra-layer hopping integrals between $d_{xz}$-like and $d_{yz}$-like orbitals on each layer of the honeycomb lattice are constructed via the Slater-Koster formalism, which reflects coexisting $\sigma$ and $\pi$ bonds.
The basis is labeled {$\left|d_{xz}^{A1}\right\rangle$, $\left|d_{yz}^{A1}\right\rangle$, $\left|d_{xz}^{B1}\right\rangle$, $\left|d_{yz}^{B1}\right\rangle$, $\left|d_{xz}^{A2}\right\rangle$, $\left|d_{yz}^{A2}\right\rangle$, $\left|d_{xz}^{B2}\right\rangle$, $\left|d_{yz}^{B2}\right\rangle$} with $A1, B1$ and $A2, B2$ denoting the sublattices of the upper layer and the lower layer [Fig. \ref{fig:Model}(a)]. The total Hamiltonian in the reciprocal space can be written as 
	\begin{equation}
	H_{8} = H_{intra} \tau_{0} + H_{inter},
	\end{equation}
where $H_{intra}$ represents the intra-layer Hamiltonian of one layer, $H_{inter}$ represents the inter-layer Hamiltonian and the Pauli matrix $\tau_{0}$ act on the layer degree of freedom.
$H_{intra}$ is a $4 \times 4$ block and has the form of 
\begin{equation}
H_{intra}=\left(\begin{array}{cccc}
0 & 0 & h13 & h14 \\
& 0 & h23 & h24 \\
& \dagger & 0 & 0 \\
& & & 0
\end{array}\right),
\end{equation}
where the matrix elements are given as 
\begin{equation}
\begin{split}
h13&=\left(\frac{t_{\pi}}{4}+\frac{3t_{\sigma}}{4}\right)e^{i\left(\frac{k_{x}}{3}-\frac{k_{y}}{3}\right)}+\left(\frac{t_{\pi}}{4}+\frac{3t_{\sigma}}{4}\right)e^{i\left(-\frac{2\text{\ensuremath{k_{x}}}}{3}-\frac{k_{y}}{3}\right)}+t_{\pi}e^{i\left(\frac{\text{\ensuremath{k_{x}}}}{3}+\frac{2\text{\ensuremath{k_{y}}}}{3}\right)},\\
h24&=\left(\frac{3t_{\pi}}{4}+\frac{t_{\sigma}}{4}\right)e^{i\left(\frac{k_{x}}{3}-\frac{k_{y}}{3}\right)}+\left(\frac{3t_{\pi}}{4}+\frac{t_{\sigma}}{4}\right)e^{i\left(-\frac{2\text{\ensuremath{k_{x}}}}{3}-\frac{k_{y}}{3}\right)}+t_{\sigma}e^{i\left(\frac{\text{\ensuremath{k_{x}}}}{3}+\frac{2\text{\ensuremath{k_{y}}}}{3}\right)},\\
h14&=h23=\frac{\sqrt{3}}{4}(t_{\sigma}-\text{\ensuremath{t_{\pi}}})e^{i\left(-\frac{2k_{x}}{3}-\frac{\text{\ensuremath{k_{y}}}}{3}\right)}-\frac{\sqrt{3}}{4}(t_{\sigma}-\text{\ensuremath{t_{\pi}}})e^{i\left(\frac{k_{x}}{3}-\frac{\text{\ensuremath{k_{y}}}}{3}\right)}.
\end{split}
\end{equation}
$H_{inter}$ is a $8 \times 8$ block and has the form of 
\begin{equation}
H_{inter} = \left(\begin{array}{cc}
0 & H_{inter}^{u} \\
\dagger & 0
\end{array}\right),
\end{equation}
with
\begin{equation}
H_{inter}^{u} = \left(\begin{array}{cccc}
hi11 & hi12 & 0 & 0 \\
hi21 & hi22 & 0 & 0 \\
0 & 0 & hi11 & hi12 \\
0 & 0 & hi21 & hi22
\end{array}\right).
\end{equation}
The matrix elements read
\begin{equation}
\begin{split}
hi11& =  \frac{r_2}{2}e^{\frac{ik_z}{2}} +\frac{r_2}{2}e^{-\frac{i k_z}{2}},\\
hi12& = -\frac{r_1}{2}e^{\frac{ik_z}{2}} -\frac{r_1}{2}e^{-\frac{i k_z}{2}},\\
hi21& = \frac{r_1}{2}e^{\frac{ik_z}{2}} +\frac{r_1}{2}e^{-\frac{i k_z}{2}},\\ 
hi22& = \frac{r_2}{2}e^{\frac{ik_z}{2}} +\frac{r_2}{2}e^{-\frac{i k_z}{2}}.
\end{split}
\end{equation}

The generators of space group P6/mmc are $C_{2z}$, $C_{3z}$, $P$ and $C_{2xy}^{'}=\lbrace C_{2xy} \vert 0 0 \frac{1}{2} \rbrace$. The representations of these symmetry operators on the above basis are given by
\begin{equation}
\begin{split}
D(C_{3z})& = -e^{\frac{i \pi}{3} s_y} \sigma_0 \tau_0,\\
D(C_{2z})& = -s_0 \sigma_x \tau_0,\\
D(P)& = s_0 \sigma_x \tau_0,\\
D(C_{2xy}^{'})& = (-\frac{\sqrt{3}}{2} s_x+\frac{1}{2}s_z) \sigma_x \tau_x.
\end{split}
\end{equation}
The Hamiltonian satisfies the equation of $D(R)^{-1}H_{8}(k)D(R)=H(R^{-1}k)$, where $R$ represents the above four symmetry operators. Therefore, the Hamiltonian belongs to the space group P6/mmc.

We use the parameters of $t_{\sigma}=t$, $t_{\pi}=-2t$, $r_1=0.4t$, $r_2=-0.3t$, and the band structure is shown in Fig. \ref{fig:Model}(b). The number of the crossing points on $\Theta = \pi$ in the Wilson loop of a sphere enclosing the DP is one indicating the nontrivial  $\mathbb{Z}_2$ topology of the DP, as shown in the left panel of Fig. \ref{fig:Model}(c). To understand the higher-order bulk-boundary correspondence, we calculate the Wilson loop spectrum at the planes of $k_z \in(-\pi,\pi)$, with the $k_z=0$ plane shown in the right panel of Fig. \ref{fig:Model}(c). The number of the crossing points on $\Theta = 0$ and $\Theta = \pi$ in the Wilson loop is both one, which indicates the $w_2 = 1$. Each slice with a specific $k_z$ in the Brillouin zone (BZ) is a torus and can be taken as a 2D subsystem. The all $k_z$ slices except $k_z =\pm \pi$ carry nontrivial $w_2=1$. Therefore, each slice in the region of (-$\pi, \pi$) is a 2D Stiefel-Whitney insulator, which has a pair of topologically protected corner zero modes. Such zero modes from all of these nontrivial $k_z$ slices constitute the topological protected hinge states, as shown in Fig. \ref{fig:Model}(d). The hinge states are similar to the ones obtained from DFT calculation, which distributes in the entire Brillouin zone along the $k_z$.

We can induce other novel topological child phases from the $\mathbb{Z}_2$DP parent phase. A perturbation term of $m_0s_0 \sigma_0 \tau_z$ with $m_0=0.3t$ is added in the minimal TB lattice model $H_{8}$, which breaks the symmetry of $C_{2xy}^{'}$. The degeneracy at the $H$ point is thus broken, and the DP at the $H$ point is split into NLs. The band structure and the momentum distribution of the NLs are shown in Figs. \ref{fig:Model}(e) and \ref{fig:Model}(g). The number of the crossing points on $\Theta = \pi$ in the Wilson loop of a sphere enclosing the NL is one indicating the nontrivial $\mathbb{Z}_2$ topology of the NL, as shown in the left panel of Fig. \ref{fig:Model}(f). In Fig. \ref{fig:Model}(g), the red circles are $\mathbb{Z}_2$ NLs from the valence band and the conduction band, and the blue lines are NLs from the two highest valence bands or two lowest conduction bands. The two kinds of NLs link with each other. Similar to the above analysis on the DP phase, we calculate the Wilson loop spectrum at the planes of $k_z \in(-\pi,\pi)$ and find all the Wilson loops nontrivial, with the $k_z=0$ plane shown in the right panel of Fig. \ref{fig:Model}(f). Therefore, each slice in the region of (-$\pi, \pi$) is a 2D Stiefel-Whitney insulator, which has a pair of topologically protected corner zero modes. Such zero modes from all of these nontrivial $k_z$ slices also constitute the topological protected hinge states, as shown in Fig. \ref{fig:Model}(h). Such a scenario to induce $\mathbb{Z}_2$NL phase can be realized in 3D TBG with uniaxial strain applied, as demonstrated in Section \ref{uniaxial_strain_TBG}.

Tailoring the parameters of interlayer coupling can lead to another pair of DPs along the high symmetric line $\Gamma A$ in addition to the pair of DPs at the points $H/H'$, which are labeled as DP2 and DP1 respectively, as shown in Fig. \ref{fig:Model}(i) with the parameters same as those of the one pair of $\mathbb{Z}_2$DPs case but $r_1=0.9t$. The DP2 is an accidental degenerate point while the DP1 is an essential degenerate point. Both types of DPs have nontrivial $\mathbb{Z}_2$ topology, since the number of the crossing points on $\Theta = \pi$ in their Wilson loops of a sphere enclosing the DP is one, as shown in Fig. \ref{fig:Model}(j). We denote the coordinates of the $z$ component of this pair of DP2 as $\pm k_{D2}$. To comprehend the higher-order bulk-boundary correspondence, we calculate the Wilson loop spectrum at the slices of $k_{z}\in(-\pi,-k_{D2})\cup(-k_{D2},k_{D2})\cup(k_{D2},\pi)$. We find $w_2$ of the slices between the pair of DP2, i.e. $k_{z}\in(-k_{D2},k_{D2})$, is trivial with the $k_z=0$ plane shown in the left panel of Fig. \ref{fig:Model}(k) since the number of the crossing points on $\Theta = 0$ and $\Theta = \pi$ in the Wilson loop are both zero indicating the $w_2 = 0$. However, the $w_2$ of the slices of $k_{z}\in(-\pi,-k_{D2})\cup(k_{D2},\pi)$, is nontrivial with the $k_z=0.6 \pi$ plane shown in right panel of Fig. \ref{fig:Model}(k) since the number of the crossing points on $\Theta = 0$ and $\Theta = \pi$ are both one indicating the $w_2 = 1$. Therefore, the hinge states are split into two pieces with the length tunable by the parameters of interlayer coupling as shown in Fig. \ref{fig:Model}(l).

\section{Full $p_z$-orbital Slater-Koster tight-binding model of 3D TBG}
	
	\begin{figure}[h]
		\begin{center}
			\includegraphics[width=0.9\linewidth]{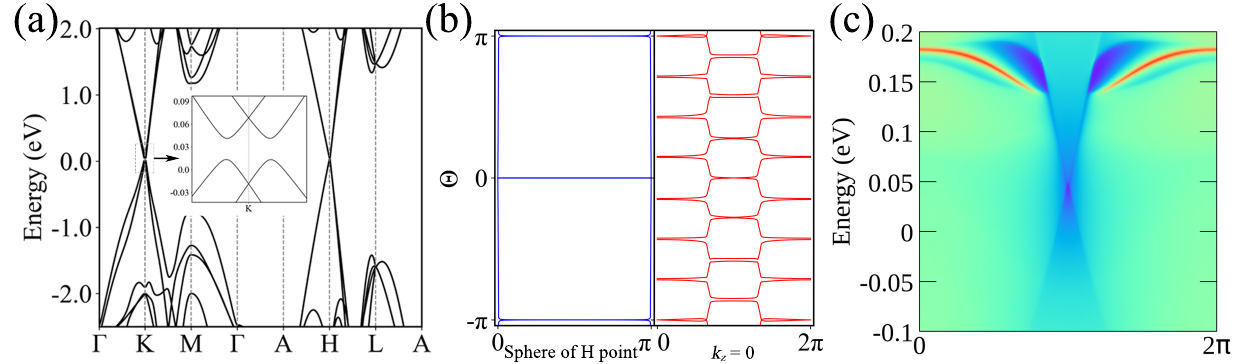} 
		\end{center} 
		\caption{Band structure, Wilson loop, and hinge states of full $p_z$-orbital Slater-Koster TB model. (a) The band structure of the TB model. The inset shows the magnified view near the $K$ point. (b) Wilson loops of a sphere enclosing the $H$ point and the torus of $k_z=0$. (c) Hinge states of the full $p_z$-orbital Slater-Koster TB model along the $k_z$ direction with 25 unit cells at the $Y$ direction.} \label{fig:TB}
	\end{figure}	
	
We also construct a TB model with full $p_z$-orbitals of the 3D TBG, similar to the 2D TBG \cite{MoonPRB2013}. The Hamiltonian can be written as
	\begin{equation}
H_{SK}=\sum_{i , j} t_{i, j} c_{i}^{\dagger} c_{j} + h.c.,
	\end{equation}
where $i$, $j$ stand for the atomic sites, and $t_{i, j}$ represent the transfer integrals between site $i$ and site $j$, which has the form of 
\begin{align}
t(i, j) &=V_{p p \pi}\left[1-\left(\frac{\mathbf{d} \cdot \mathbf{e}_{z}}{d}\right)^{2}\right]+V_{p p \sigma}\left(\frac{\mathbf{d} \cdot \mathbf{e}_{z}}{d}\right)^{2}, \\
V_{p p \pi} &=V_{p p \pi}^{0} \exp \left(-\frac{d-a_{0}}{\delta_{0}}\right), \\
V_{p p \sigma} &=V_{p p \sigma}^{0} \exp \left(-\frac{d-d_{0}}{\delta_{0}}\right).
\end{align}
$\mathbf{d}$ and $d$ are the vector and distance of the site $i$ to $j$, respectively. $a_0$ = 1.42 {\AA} is the distance of the neighboring sublattices on the monolayer graphene, and $d_0 = 3.35$ {\AA} is the interlayer distance.  $V_{p p \pi}^{0}$ and $V_{p p \sigma}^{0}$ are transfer integrals of the intralayer and interlayer, respectively, and we take $V_{p p \pi}^{0} = -2.7$ eV and $V_{p p \sigma}^{0} = 0.58$ eV here. $\delta_{0}$ is the decay length of the transfer integral and is chosen as 0.319 $a_0$. The transfer integrals for $d > 4a_0$ are too small to be neglected, and the interlayer hopping between only the adjacent layers is considered.

The low-energy band structure of the TB model has a good agreement with the DFT results, which both are gaped at the $K$ points and four-fold degenerate at the $H$ points [Fig. \ref{fig:TB}(a)]. The gap size of the $K$ points is related to the interlayer coupling. The gap size increases with the increasing of the transfer integrals of the interlayer and is about 30 meV when $V_{p p \sigma}^{0} = 0.58$ eV. The Dirac points at $H$ point of the TB model also carry a nontrivial $\mathbb{Z}_2$ monopole charge as plotted in the left panel of Fig. \ref{fig:TB}(b). The $w_2$ of $k_z$ slices from $-\pi$ to $\pi$ are nontrivial with the $k_z=0$ plane shown in the right panel of Fig. \ref{fig:TB}(b), leading to a pair of hinge Fermi arcs as shown in Fig. \ref{fig:TB}(c).

\section{3D TBG under pressure}

	\begin{figure}[h]
		\begin{center}
			\includegraphics[width=1\linewidth]{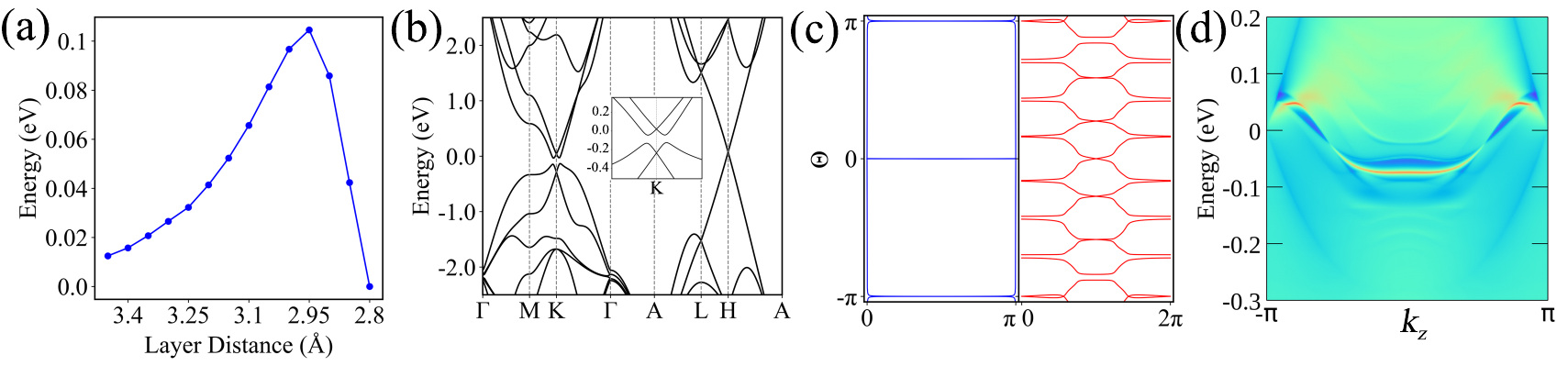} 
		\end{center} 
		\caption{Band gap, band structure, Wilson loop, and hinge states of 3D TBG under pressure. (a)The layer distance dependence of band gap at $k_z = 0$ of 3D TBG from DFT calculation. (b) The band structure of 3D TBG when the layer distance is 3.05 \AA . (c) Wilson loops of a sphere enclosing the $H$ point and the torus of $k_z=0$ when the layer distance is 3.05 \AA . (d) Hinge states of 3D TBG along the $k_z$ direction with 8 unit cells at the $Y$ direction when the layer distance is 3.05 \AA .} \label{fig:pressure}
	\end{figure}
	
	 The 3D TBG has a layered structure and layers are bounded with weak van der Waals forces. Therefore, the layer distance can be significantly changed by the pressure. We calculate the band gap at the plane of $k_z=0$ with the different layer distances corresponding to different pressures, as shown in Fig. \ref{fig:pressure}(a). The band gap can be up to 0.1 eV at the layer distance of 2.95 \AA . Figure \ref{fig:pressure}(b) shows the band structures at the layer distance of 3.05 \AA . A sizable band gap opens near the $K$ point. The Wilson loop shows that the nontrivial topology of the 3D TBG at the layer distance of 3.05 \AA\ does not change [Fig. \ref{fig:pressure} (c)], and the hinge Fermi arcs are more obvious [Fig. \ref{fig:pressure}(d)].

	\section{Uniaxial strain in 3D TBG} 	
	\label{uniaxial_strain_TBG}
			\begin{figure}[h]
		\begin{center}
			\includegraphics[width=1\linewidth]{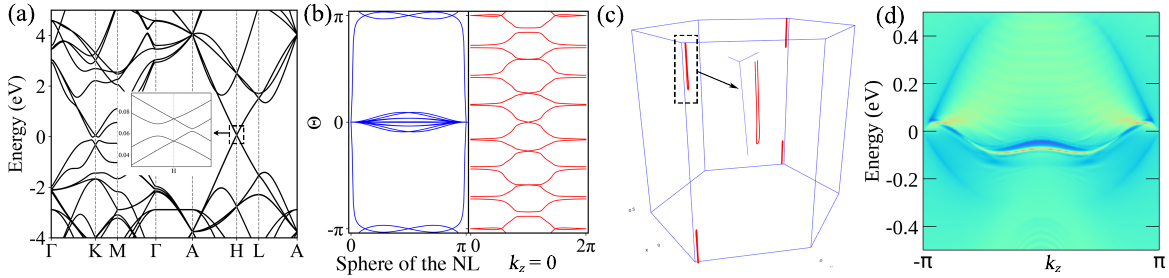} 
		\end{center} 
		\caption{The $\mathbb{Z}_2$NL phase in 3D TBG with a uniaxial strain. (a) The band structure of 3D TBG with -0.14\% uniaxial strain applied along $\mathbf{a_1}$ direction. (b) Wilson loops of a sphere enclosing a NL and the torus of $k_z=0$ of 3D TBG with a uniaxial strain. (c) Momentum distribution of the $Z_2$NLs in 3D TBG with a uniaxial strain. (d) Hinge states of 3D TBG with a uniaxial strain along the $k_z$ direction with 8 unit cells at the $Y$ direction.} \label{fig:uniaxial}
	\end{figure}
	
	The uniaxial strain applied along $x$ or $y$ direction reduces the symmetry from P6/mmc to P2/m but preserves $\mathcal{P}\mathcal{T}$. Therefore, The $\mathbb{Z}_2$DPs are expanded into $\mathbb{Z}_2$NLs as shown in Fig. \ref{fig:uniaxial}. The momentum distribution of the NLs is shown in Fig. \ref{fig:uniaxial}(c), which are located near the high symmetric line $KH$. The topology of the NL is confirmed by the Wilson loop of a sphere enclosing the entire NL, which has a nontrivial $w_2$ as shown in the left panel of Fig. \ref{fig:uniaxial}(b). The hinge states between the NLs are shown in Fig. \ref{fig:uniaxial}(d).

	\section{Green function method in the calculation of hinge states} 
\begin{figure}
		\begin{center}
			\includegraphics[width=0.9\linewidth]{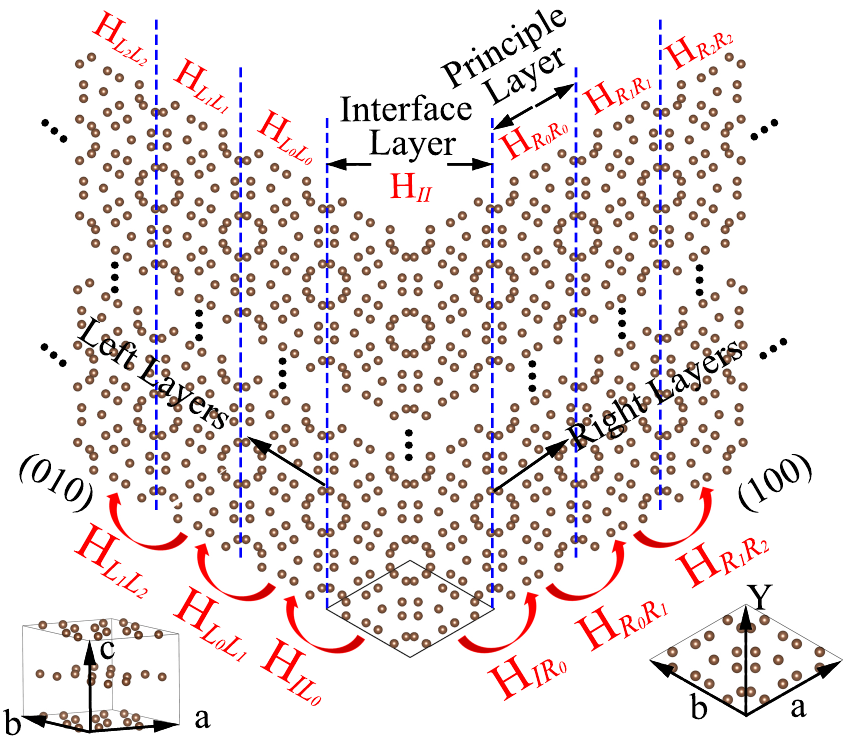} 
		\end{center} 
		\caption{The geometry of bi-semi-infinite open boundary of 3D TBG. The geometry is divided into an interface layer and principle layers (right layers and left layers). The hopping matrix between each principle layer and interface layer is given in the figure.} \label{fig:hingestructure}
	\end{figure}

The hinge states of the model and materials in the main text are calculated using recursive Green function method \cite{sancho1985highly, yueSymmetryenforcedChiralHinge2019}. We designed a bi-semi-infinite open boundary geometry as shown in Fig. \ref{fig:hingestructure}. The geometry is semi-infinite along the $\mathbf{a}$ and $\mathbf{b}$ directions, finite along the $\mathbf{Y}$ direction, and periodic along the $\mathbf{c}$ direction. The geometry is divided into an interface layer and principle layers (left layers and right layers). The Hamiltonian can be expressed as
\begin{widetext}
\begin{equation}
H=\left(\begin{array}{llllllllll}
H_{I I} & H_{I L_{0}} & H_{I R_{0}} & 0 & \cdots & & & & & \\
H_{I L_{0}}^{\dagger} & H_{L_{0} L_{0}} & 0 & H_{L_{0} L_{1}} & 0 & \cdots & & & & \\
H_{I R_{0}}^{\dagger} & 0 & H_{R_{0}R_0} & 0 & H_{R_{0} R_{1}} & 0 & \cdots & & & \\
0 & H_{L_{0} L_{1}}^{\dagger} & 0 & H_{L_{1} L_{1}} & 0 & H_{L_{1} L_{2}} & 0 & \cdots & & \\
\vdots & 0 & H_{R_{0} R_{1}^{-}}^{\dagger} & 0 & H_{R_{1} R_{1}} & 0 & H_{R_{1} R_{2}} & 0 & \cdots & \\
& \vdots & 0 & H_{L_{1} L_{2}}^{\dagger} & 0 & H_{L_{2} L_{2}} & 0 & H_{L_{2} L_{3}} & 0 & \cdots \\
& & \vdots & 0 & H_{R_{1} R_{2}}^{\dagger} & 0 & H_{R_{2} R_{2}} & 0 & H_{R_{2} R_{3}} & \ddots \\
& & & \vdots & \ddots & \ddots & \ddots & \ddots & \ddots & \ddots
\end{array}\right).
\end{equation} 
\end{widetext} \label{Eq:S1}
The diagonal block $H_{L_{i} L_{i}} = H_{L_{0} L_{0}}$, $H_{R_{i}R_i} = H_{R_{0}R_0}$, and the off-diagonal block $H_{L_{i} L_{i+1}} = H_{L_{0} L_{1}}$, $H_{R_{i}R_{i+1}} = H_{R_{0}R_1}$.  Each block of $H$ is shown in Fig. \ref{fig:hingestructure}. The TB Hamiltonian is obtained by Wannier function interpolation.

For simplification, the $H_{L}$ denotes the Hamiltonian of left layers
\begin{equation}
H_{L} \equiv\left(\begin{array}{lllll}
H_{L_{0} L_{0}} & H_{L_{0} L_{1}} & 0 & & \\
H_{L_{0} L_{1}}^{\dagger} & H_{L_{0} L_{0}} & H_{L_{0} L_{1}} & 0 & \\
0 & H_{L_{0} L_{1}}^{\dagger} & H_{L_{0} L_{0}} & H_{L_{0} L_{1}} & \ddots \\
& \ddots & \ddots & \ddots & \ddots
\end{array}\right),
\end{equation}
and the $H_{R}$ denotes the Hamiltonian of right layers
\begin{equation}
H_{R} \equiv\left(\begin{array}{lllll}
H_{R_{0} R_{0}} & H_{R_{0} R_{1}} & 0 & & \\
H_{R_{0} R_{1}}^{\dagger} & H_{R_{0} R_{0}} & H_{R_{0} R_{1}} & & \\
0 & H_{R_{0} R_{1}}^{\dagger} & H_{R_{0} R_{0}} & H_{R_{0} R_{1}} & \ddots \\
& \ddots & \ddots & \ddots & \ddots
\end{array}\right).
\end{equation}
The $H_{IP}$ denote the hoppings between interface layer $H_{II}$ and left layers, right layers:
\begin{equation}
H_{I P} \equiv\left(\begin{array}{llll}
H_{I L_{0}} & H_{I R_{0}} & 0 & \cdots
\end{array}\right).
\end{equation}
Therefore, we have
\begin{equation}
H \equiv\left(\begin{array}{ll}
H_{I I} & H_{I P} \\
H_{I P}^{\dagger} & H_{P P}
\end{array}\right),
\end{equation}
where 
\begin{equation}
H_{PP} \equiv H_{L} \tilde{\oplus} H_{R},
\end{equation}
since the left layers and right layers are decoupled from each other.

The total Green function $G$ meets
\begin{equation}
(\omega+i \eta-H) G=1,
\end{equation}
with 
\begin{equation}
G \equiv\left(\begin{array}{ll}
G_{I I} & G_{I P} \\
G_{I P}^{\dagger} & G_{PP}
\end{array}\right).
\end{equation}
The interface layer Green function has the form
\begin{equation}
G_{I I}\left(k_{z}, \omega+i \eta\right)=\left[(\omega+i \eta)-H_{I I}-\Sigma_{R}\right]^{-1},
\end{equation}
where the self-energy $\Sigma_{R}$ reads
\begin{equation}
\Sigma_{R}=H_{I P} g_{P P} H_{I P}^{\dagger} \approx H_{I L_{0}} g_{L_{0} L_{0}} H_{I R_{0}}^{\dagger}+H_{I R_0} g_{R_{0} R_{0}} H_{I R_{0}}^{\dagger}.
\end{equation}
$g_{L_{0} L_{0}}$ and $g_{R_{0} R_{0}}$ are the so-called surface Green functions of left layers and right layers, respectively, and can be solved by the standard recursive method.
The imaginary part of the interface layer Green function is the spectral function
\begin{equation}
A_{I I}\left(k_{z}, \omega\right)=-\frac{1}{\pi} \mathfrak{\Im} G_{I I}\left(k_{z}, \omega\right).
\end{equation}
The trace of $A_{II}$ at the lower hinge gives the spectrum of the hinge states.

\section{Hinge states of 3D TBG} 
	\begin{figure}[b]
		\begin{center}
			\includegraphics[width=0.9\linewidth]{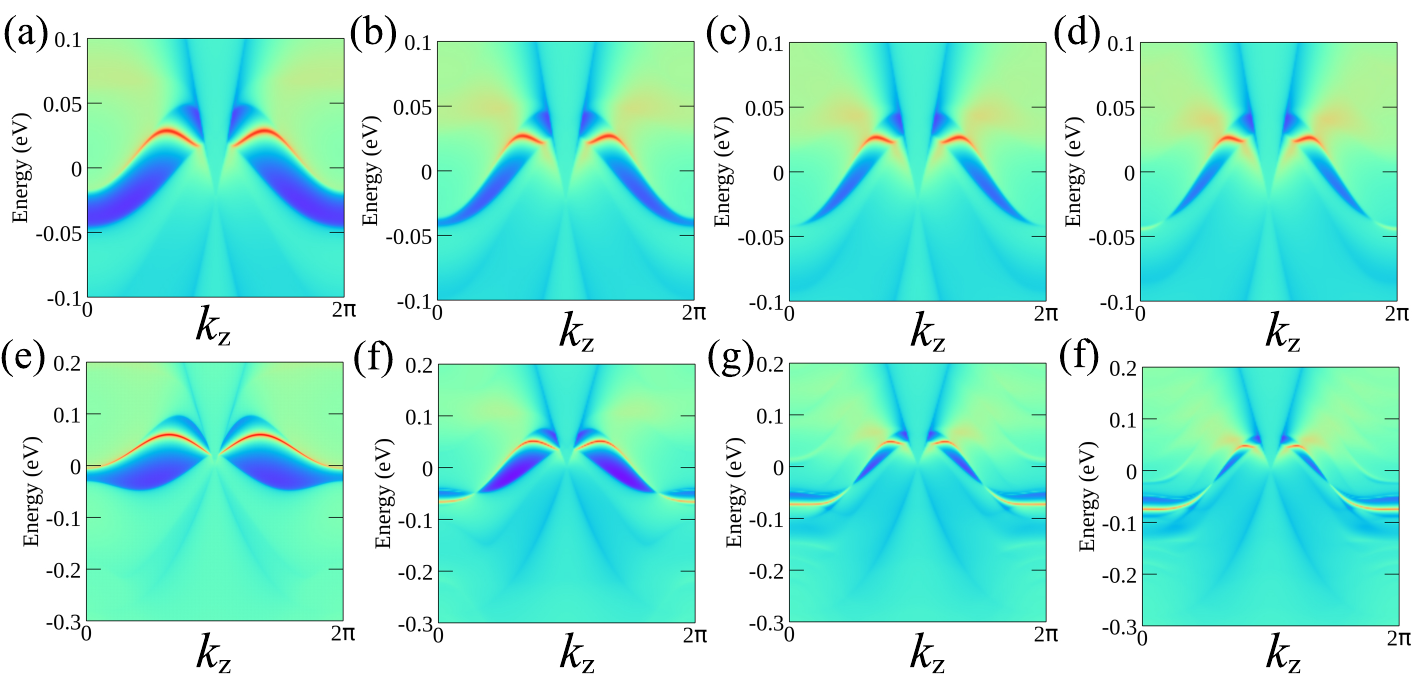}
		\end{center} 
		\caption{Hinge states with different size of $Y$ direction of 3D TBG and 3D TBG in pressure. (a-d) The hinge states of 3D TBG at the $k_z$ direction with 20, 30, 35, and 40 unit cells along $Y$ direction, respectively. (e-h) The hinge states of 3D TBG with layer distance of 3.05 {\AA} along the $k_z$ direction with 8, 15, 30, and 40 unit cells along $Y$ direction, respectively.} \label{fig:hinge}
	\end{figure}
	
\begin{figure}[h]
		\begin{center}
			\includegraphics[width=0.9\linewidth]{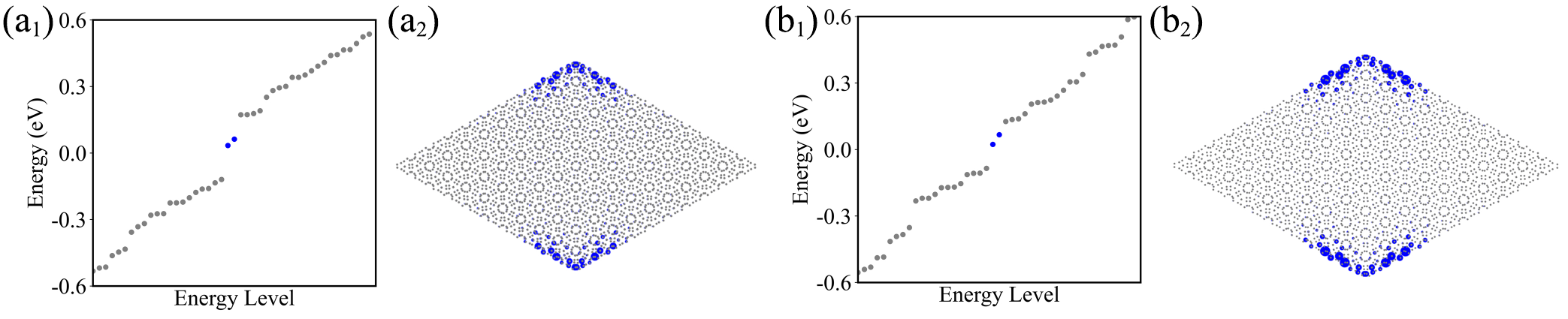} 
		\end{center} 
		\caption{Hinge states of finite-size 3D TBG along both $\mathbf{a}$ and $\mathbf{b}$ directions. (a1) Energy spectrum of a rhombus finite-size  3D TBG flake of $k_z=0$ along both $\mathbf{a}$ and $\mathbf{b}$ directions with 100 unit cells calculated by the Wannier interpolation method. The blue dots represent the hinge states, whose real-space distribution of the two restricted dimensions are plotted in (a2). (b1), (b2) The same as (a1) and (a2) but with $k_z=\pi/4$.} \label{fig:hingeslab}
	\end{figure}

Figure \ref{fig:hinge} shows the hinge states at different sizes of the Y direction using the above recursive interface Green function method. Figure \ref{fig:hinge}(a-d) show the 3D TBG in the normal condition at the size of 20, 30, 35, and 40 unit cells along $Y$ direction, respectively. We also calculate the hinge states of 3D TBG under pressure with a larger band gap at $k_z=0$. In addition, we calculate the hinge states of 3D TBG using a slab geometry which is finite along both $\mathbf{a}$ and $\mathbf{b}$ directions and periodic along $\mathbf{c}$ direction.  Figures \ref{fig:hingeslab}(a, b) show the hinge states at the $k_z = 0$ and $k_z = \pi/4$, respectively. The blue dots mark the hinge states and the corresponding real space distributions are shown in the Fig. \ref{fig:hingeslab}(c) and Fig. \ref{fig:hingeslab}(d), respectively.

\section{Effective models of $\mathbb{Z}_2$DPs in all other possible space groups}	
Similar to developing the effective model of $\mathbb{Z}_2$DPs in Space Group of No. 192 in Section \ref{Section_Heff_192} we also construct the effective models of $\mathbb{Z}_2$DPs in all other 11 possible space groups with the help of MagneticKP \citep{zhangMagneticKP2022}.  In the following, we give the space group numbers, location of the $\mathbb{Z}_2$DPs, corresponding generating elements, and corepresentation of the little group, $\mathbf{k} \cdot \mathbf{p}$ Hamiltonian of the $\mathbb{Z}_2$DPs with the analytical and numerical energy spectra, and Wilson loops, one by one. The notations of the symmetry representation can be found in Refs. \cite{bradley2010mathematical, liuMSG2022}.

The essential DP at the high-symmetric $W$ point $\left(\frac{3}{4}, \frac{\overline{1}}{4}, \frac{\overline{1}}{4}\right)$ of space group Ibca (No. 73) is protected by $C_{2 z}, C_{2 y}, \mathcal{P} \mathcal{T}$ symmetries, and the corepresentations of the above symmetries can be written as 
	\begin{equation}
D(C_{2 z}) = \Gamma_{0,1}, D(C_{2 y}) = \Gamma_{0,3}, D(\mathcal{P} \mathcal{T}) = -\Gamma_{2,2}, 
	\end{equation}
	Constrained by these symmetries, the Hamiltonian expanded to the first order of $\mathbf{k}$ is 
	\begin{equation}
	H(k) = \sum_{i=1}^3\left(\alpha_{i, 1} \Gamma_{i, 2} k_x+\alpha_{i, 2} \Gamma_{i, 3} k_y+\alpha_{i, 3} \Gamma_{i, 1} k_z\right),
    \end{equation}
    where $\alpha_{i, j}$ are parameters and $\Gamma_{i, j} = \sigma_i \otimes \sigma_j$. With all parameters equal to 1, the eigenvalues are 
    \begin{equation}
      E = \pm \sqrt{3} \sqrt{k_x^2+k_y^2+k_z^2}.  
    \end{equation}

\begin{figure}[h]
		\begin{center}
			\includegraphics[width=0.9\linewidth]{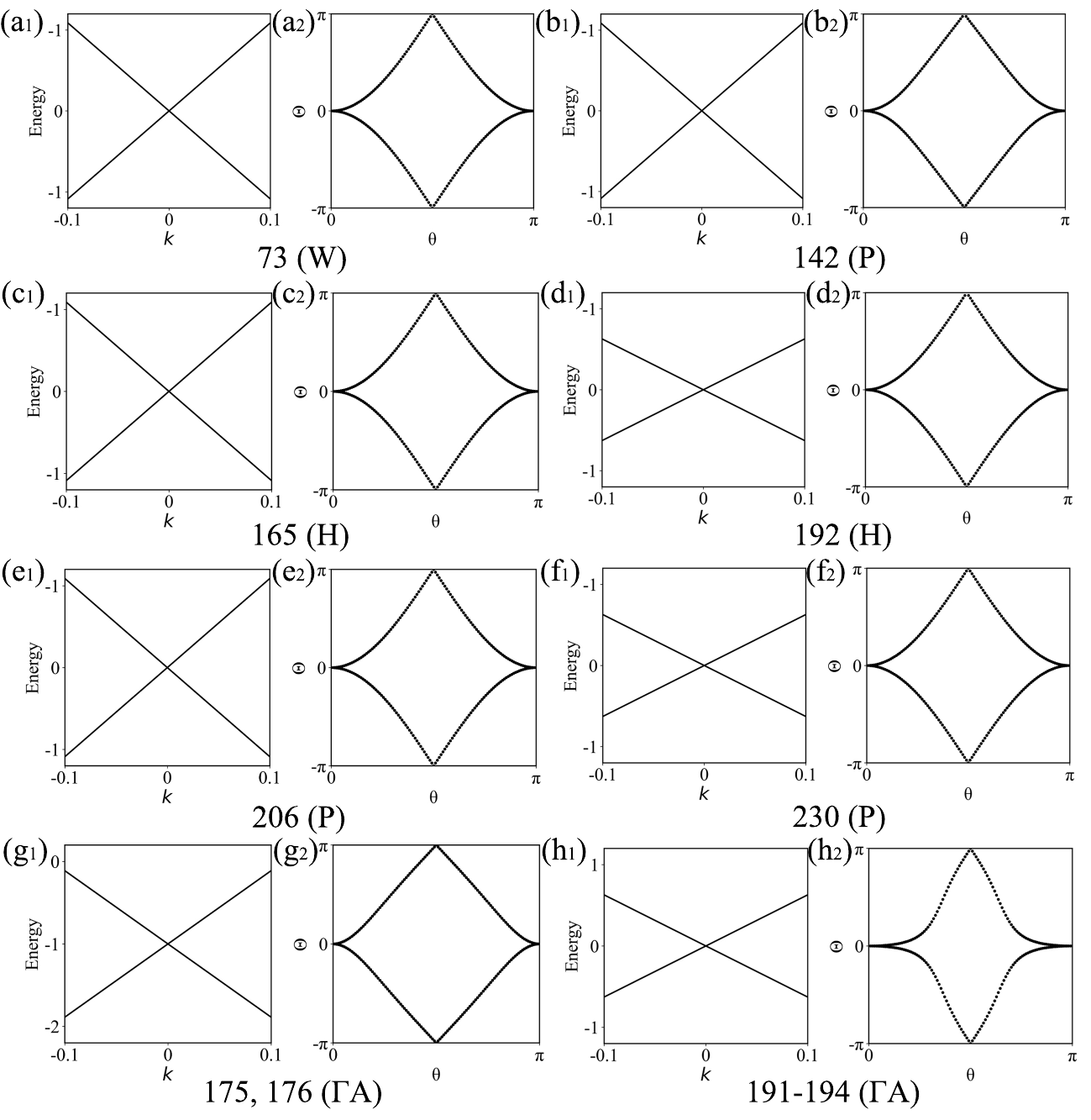} 
		\end{center} 
		\caption{Band structures and Wilson loops of the low-energy effective models of $\mathbf{Z}_2$DPs in all possible space groups (12 space groups). The band structures and Wilson loop enclosing the $\mathbf{Z}_2$DPs of the effective Hamiltonian around the high-symmetric point $W$ of space group No. 73 (a$_1$ and a$_2$), the high-symmetric point $P$  of space group No. 142 (b$_1$, b$_2$), the high-symmetric point $H$ of space group No. 165 (c$_1$ and c$_2$), the high-symmetric point $H$  of space group No. 192 (d$_1$ and d$_2$), the high-symmetric point $P$  of space group No. 206 (e$_1$ and e$_2$) and the high-symmetric point $P$  of space group No. 230 (f$_1$ and f$_2$),  and along the high-symmetric line $\Gamma A$  of space group No. 175 and 176 (g$_1$ and g$_2$) and the high-symmetric line $\Gamma A$  of space groups Nos. 191-194 (h$_1$ and h$_2$).} \label{fig:band_wcc}
	\end{figure}

   The band structure and Wilson loop enclosing the DP are shown in the Figs. \ref{fig:band_wcc}($a_1$, $a_2$), which demonstrate the nontrivial $\mathbb{Z}_2$ monopole charge.

    The essential DP at the high-symmetric $P$ point $\left(\frac{1}{4}, \frac{1}{4}, \frac{1}{4}\right)$ of space group I4$_1$/acd (No. 142) is protected by $S_{4 z}^{-}, C_{2 y}, \mathcal{P} \mathcal{T}$ symmetries, and the corepresentations of the above symmetries can be written as 
	\begin{equation}
D(S_{4 z}^{-}) = \left(\frac{1}{2}-\frac{i}{2}\right)\left(\Gamma_{0,3}+i \Gamma_{3,0}\right), D(C_{2 y}) = \Gamma_{3,1}, D(\mathcal{P} \mathcal{T}) = \Gamma_{1,0}.
	\end{equation}
	Constrained by these symmetries, the Hamiltonian expanded to the first order of $\mathbf{k}$ reads 
	\begin{equation}
	H(k) = \alpha_1\left(\Gamma_{3,2} k_x+\Gamma_{0,1} k_y\right)+\alpha_2\left(\Gamma_{2,0} k_x-\Gamma_{1,3} k_y\right) +\alpha_3\left(\Gamma_{1,0} k_x+\Gamma_{2,3} k_y\right)+k_z\left(\alpha_3 \Gamma_{1,1}-\alpha_4 \Gamma_{2,1}\right),
    \end{equation}
    where $\alpha_{i}$ are parameters and $\Gamma_{i, j} = \sigma_i \otimes \sigma_j$. With all parameters equal to 1, the eigenvalues are 
    \begin{equation}
E = \pm \sqrt{3 k_x^2+3 k_y^2+2 k_z^2-2 \sqrt{2 k_z^2 \left(k_x^2+k_y^2\right)+8 k_x^2 k_y^2}}.  
    \end{equation}
The band structure and Wilson loop enclosing the DP are shown in the Figs. \ref{fig:band_wcc}($b_1$, $b_2$), which demonstrate the nontrivial $\mathbb{Z}_2$ monopole charge.

    The essential DP at the high-symmetric $H$ point $\left(\frac{\overline{1}}{3}, \frac{2}{3}, \frac{1}{2}\right)$ of space group P$\overline{3}c1$ (No. 165) is protected by $C_3^{+}, C_{21}^{\prime \prime}, \mathcal{P} \mathcal{T}$ symmetries, and the corepresentations of the above symmetries can be written as 
	\begin{equation}
D(C_3^{+}) = \frac{\Gamma_{0,0}-i \sqrt{3} \Gamma_{0,2}}{2}, D(C_{21}^{\prime \prime}) = \Gamma_{0,3}, D(\mathcal{P} \mathcal{T}) = -\Gamma_{2,2}.
	\end{equation}
	Constrained by these symmetries, the Hamiltonian expanded to the first order of $\mathbf{k}$ is 
	\begin{equation}
	H(k) = \sum_{i=1}^3\left[\alpha_{i, 1}\left(\Gamma_{i, 1} k_x-\Gamma_{i, 3} k_y\right)+\alpha_{i, 2} k_z \Gamma_{i, 2}\right],
    \end{equation}
    where $\alpha_{i, j}$ are parameters and $\Gamma_{i, j} = \sigma_i \otimes \sigma_j$. The eigenvalues of the Hamiltonian are 
    
       \begin{equation}
    E = \pm \sqrt{k_z^2(\alpha_{1, 1}^2+\alpha_{1, 2}^2+\alpha_{1, 3}^2)+(k_x^2+k_y^2)(\alpha_{2, 1}^2+\alpha_{2, 2}^2+\alpha_{2, 3}^2) \pm 2m}, 
      \end{equation}
 where $m=\sqrt{(k_x^2+k_y^2)k_z^2((\alpha_{1, 2} \alpha_{2, 1}-\alpha_{1, 1} \alpha_{2, 2})^2 - (\alpha_{1, 3}
      \alpha_{2, 1} -  \alpha_{1, 1} \alpha_{3, 2})^2 - (\alpha_{1, 3}  \alpha_{2, 2} - \alpha_{1, 2} \alpha_{3, 2})^2)}$. With all parameters equal to 1,  the band structure and Wilson loop enclosing the DP are shown in the Figs. \ref{fig:band_wcc}($c_1$, $c_2$), which demonstrate the nontrivial $\mathbb{Z}_2$ monopole charge.

    The essential DP at the high-symmetric $P$ point $\left(\frac{1}{4}, \frac{1}{4}, \frac{1}{4}\right)$ of space group Ia$\overline{3}$ (No. 206) is protected by $C_{31}^{+}, C_{2 x}, C_{2 z}, \mathcal{P} \mathcal{T}$ symmetries, and the corepresentations of the above symmetries can be written as 
	\begin{equation}
	\begin{split}
D(C_{31}^{+}) = \frac{\sqrt{3} \Gamma_{0,3}+i \Gamma_{0,0}}{2}, D(C_{2 x}) = -\frac{\Gamma_{0,1}+\Gamma_{0,2}+\Gamma_{0,3}}{\sqrt{3}},  \\
D(C_{2 z})=\frac{-(\sqrt{3}-3) \Gamma_{0,1}-(3+\sqrt{3}) \Gamma_{0,2}+2 \sqrt{3} \Gamma_{0,3}}{6},D(\mathcal{P} \mathcal{T}) = -\Gamma_{2,2}.
    \end{split}	
	\end{equation}
	Constrained by these symmetries, the Hamiltonian expanded to the first order of $\mathbf{k}$ is 
	\begin{equation}
	H(k) = \sum_{i=1}^3 \alpha_{i, 1}[\Gamma_{i, 3} q_c+((-1)^{5 / 12} q_{-} \Gamma_{i,+}+ h.c.)],
    \end{equation}
    where $\alpha_{i, j}$ are parameters, $\Gamma_{i, j} = \sigma_i \otimes \sigma_j$ ,$\Gamma_{i, +} = \sigma_i \otimes (\sigma_x+i\sigma_y)/2$, $\{q_a, q_b, q_c\} = \{k_x + k_y + k_z, -2 k_x + k_y + \sqrt{3} k_y + k_z - \sqrt{3} k_z,  2 k_x+\left(\sqrt{3}-1\right) k_y-\left(\sqrt{3}+1\right) k_z\}$ and $q_{-}=(q_a-iq_b)/2$. The eigenvalues of the Hamiltonian are 
    \begin{equation}
    \begin{split}
E = \pm \sqrt{(\alpha_{1, 1}^2+\alpha_{2, 1}^2+\alpha_{3, 1}^2)(q_x^{2}+q_y^{2}+q_z^{2})}.
	\end{split}
    \end{equation}
    With all parameters equal to 1, the band structure and Wilson loop enclosing the DP are shown in the Figs. \ref{fig:band_wcc}($e_1$, $e_2$), which demonstrate the nontrivial $\mathbb{Z}_2$ monopole charge.

    The essential DP at the high-symmetric $P$ point $\left(\frac{1}{4}, \frac{1}{4}, \frac{1}{4}\right)$ of space group Ia$\overline{3}$d (No. 230) is protected by $C_{32}^{+}, C_{2 y}, C_{2 x}, S_{4 x}^{+}, \mathcal{P} \mathcal{T}$ symmetries, and the corepresentations of the above symmetries can be written as 
	\begin{equation}
	\begin{split}
D(C_{32}^{+}) = \frac{\sqrt{3} \Gamma_{3,3}-i \Gamma_{0,0}}{2}, D(C_{2 y}) = i (\frac{\Gamma_{0,2}-\Gamma_{3,1}-\Gamma_{3,3}}{i \sqrt{3}}), D(C_{2 x}) =\frac{(3+\sqrt{3}) \Gamma_{0,2}-(\sqrt{3}-3) \Gamma_{3,1}+2 \sqrt{3} \Gamma_{3,3}}{6},  \\
D(S_{4 x}^{+}) = \frac{(1+i)\left[-i(\sqrt{3}-3) \Gamma_{0,1}+2 i \sqrt{3} \Gamma_{0,3}+6 \Gamma_{3,0}+i(3+\sqrt{3}) \Gamma_{3,2}\right]}{12}, D(\mathcal{P} \mathcal{T}) = \Gamma_{1,0}.
    \end{split}
	\end{equation}
	Constrained by these symmetries, the Hamiltonian expanded to the first order of $\mathbf{k}$ is 
	\begin{equation}
	H(k) = \alpha_1\left(q_a^{\prime} \Gamma_{+, 0}-i q_b^{\prime} \Gamma_{+, 3}+i q_c^{\prime} \Gamma_{+, 1}\right)+h.c.,
    \end{equation}
    where $\alpha_{i}$ are parameters and $\Gamma_{+, i} = (\sigma_x+i\sigma_y)/2 \otimes \sigma_j$, $\{q_a^{\prime}, q_b^{\prime}, q_c^{\prime}\}=\{(1+\sqrt{3}) k_x-2 k_y+(\sqrt{3}-1) k_z,(1-\sqrt{3}) k_x-2 k_y-(1+\sqrt{3}) k_z,-2 k_x-2 k_y+2 k_z\}$. The eigenvalues of the Hamiltonian are 
    \begin{equation}
E = \pm 2 \left|\alpha_1 \right|( \sqrt{q_a^{\prime2}+q_b^{\prime2}+q_c^{\prime2}}).
    \end{equation}    
With all parameters equal to 1, the band structure and Wilson loop enclosing the DP are shown in the Figs. \ref{fig:band_wcc}($f_1$, $f_2$), which demonstrate the nontrivial $\mathbb{Z}_2$ monopole charge.

    The accidental DPs along the high-symmetric lines $\Gamma A$ $\left( 0,  0,  \mu   \right)$ of space groups P6/m and P6$_3$/m (No. 175 and No. 176) are protected by $C_6^{+}, \mathcal{P} \mathcal{T}$ symmetries, and the corepresentations of the above symmetries can be written as 
	\begin{equation}
D(C_6^{+}) = \frac{\Gamma_{3,0}+i \sqrt{3} \Gamma_{0,3}}{2}, D(\mathcal{P} \mathcal{T}) = \Gamma_{0,1}.
	\end{equation}
	Constrained by these symmetries, the Hamiltonian expanded to the first order of $\mathbf{k}$ is 
	\begin{equation}
	H(k) = \Gamma_{1,0}\left(\alpha_3 k_x+\alpha_4 k_y\right)+\Gamma_{2,3}\left(\alpha_3 k_y-\alpha_4 k_x\right) +\Gamma_{0,0}\alpha_2 k_z+\Gamma_{3,0}\left(\alpha_5+\alpha_6 k_z\right),
    \end{equation}
    where $\alpha_{i}$ are parameters and $\Gamma_{i, j} = \sigma_i \otimes \sigma_j$. The eigenvalues of the Hamiltonian are
    \begin{equation}
     E = \alpha_2 k_z \pm \sqrt{(\alpha_3^2+\alpha_4^2)(k_x^2+k_y^2)+(\alpha_5+\alpha_6 k_z)^2}.
     \end{equation}
      With all parameters equal to 1,  the band structure and Wilson loop enclosing the DP are shown in the Figs. \ref{fig:band_wcc}($g_1$, $g_2$), which demonstrate the nontrivial $\mathbb{Z}_2$ monopole charge.

    The accidental DPs along the high-symmetric lines $\Gamma A$ $\left( 0,  0,  \mu   \right)$ of space groups P6/mmm, P6/mcc, P6$_3$/mcm and P6$_3$/mmc (Nos. 191-194) are protected by $C_6^{+}, \sigma_{v 1}, \mathcal{P} \mathcal{T}$ symmetries, and the corepresentations of the above symmetries can be written as 
	\begin{equation}
D(C_6^{+}) = \frac{\Gamma_{3,0}+i \sqrt{3} \Gamma_{0,2}}{-2}, D(\sigma_{v 1}) = \Gamma_{0,3}, D(\mathcal{P} \mathcal{T}) = -\Gamma_{0,0}.
	\end{equation}
	Constrained by these symmetries, the Hamiltonian expanded to the first order of $\mathbf{k}$ is 
	\begin{equation}
	H(k) = \sum_{i=0,3} \Gamma_{i, 0} \alpha_{i, 0} k_z+\alpha_2\left(\Gamma_{1,0} k_x+\Gamma_{2,2} k_y\right),
    \end{equation}
    where $\alpha_{i, 0}$ and $\alpha_2$ are parameters and $\Gamma_{i, j} = \sigma_i \otimes \sigma_j$. The eigenvalues of the Hamiltonian are 
    \begin{equation}
	E = \alpha_{0, 0} k_z \pm \sqrt{\alpha_2^2(k_x^2+k_y^2)+2\alpha_2 (\alpha_{1, 0}k_x \pm \alpha_{2, 0}k_y)k_z+(\alpha_{1, 0}^2+\alpha_{2, 0}^2+\alpha_{3, 0}^2)k_z^2}.
    \end{equation}
     With all parameters equal to 1, the band structure and Wilson loop enclosing the DP are shown in the Figs. \ref{fig:band_wcc}($h_1$, $h_2$), which demonstrate the nontrivial $\mathbb{Z}_2$ monopole charge.

\section{$\mathbb{Z}_2$DPs in other candidate materials}	
\begin{figure}[h]
		\begin{center}
			\includegraphics[width=1\linewidth]{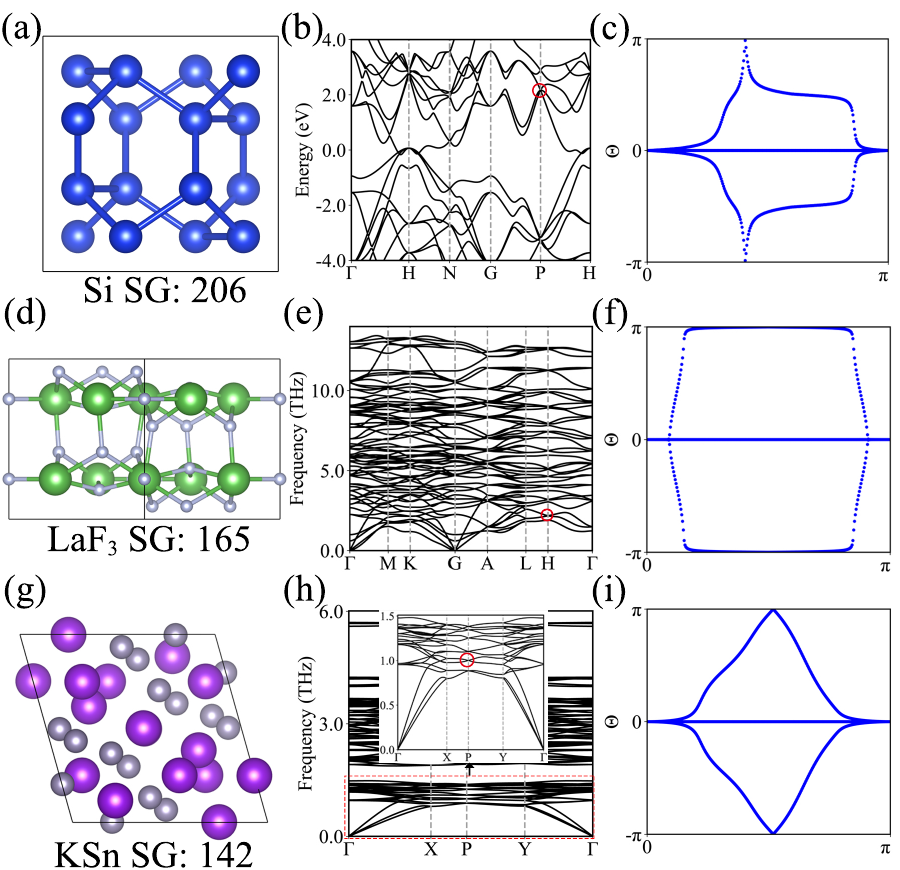}
		\end{center} 
		\caption{Geometry structures of Si (a), LaF$_3$ (d), and KSn (g) belonging to space groups 206, 165, and 142, respectively. Band structures of Si (b) and phonon spectra of LaF$_3$ (e) and KSn (h). Red circles denote the position of the $\mathbb{Z}_2$DP. Wilson loop enclosing the $\mathbb{Z}_2$DP in the red circles of Si (c), LaF$_3$ (f), and KSn (i). All the Wilson loop spectra have one crossing point on $\Theta = \pi$, indicating $w_2$ = 1.} \label{fig:othermater}
	\end{figure}

The $\mathbb{Z}_2$DP phase is not unique to ATMG. In fact, it is widely present in the band structures and phonon spectra of materials belonging to these allowable space groups in Table I in the text. For example, as shown in Fig. \ref{fig:othermater}, the $\mathbb{Z}_2$DPs are also present in the band structure at the $P$ points of Si \cite{Si206} with the space group No. 206 and ICSD No. 16569, the $H$ points of the phonon spectrum in LaF$_3$ \cite{LaF3165} with the space group No. 165 and ICSD No. 74733, and in the phonon spectrum of KSn \cite{KSn142} with the space group No. 142 and ICSD No. 104614.

\bibliographystyle{apsrev4-2}
\bibliography{reference}